\newif\ifFULL
\FULLtrue

\ifFULL
\documentclass[11pt]{article}
\usepackage{fullpage}
\usepackage{natbib}
\usepackage{hyperref}
\else

\documentclass[format=acmsmall, review=false]{acmart}

\usepackage{acm-ec-18}
\usepackage{booktabs} 
\usepackage{textcase}
\usepackage{xpatch}
\makeatletter
\xpatchcmd{\@secfont}{\MakeUppercase}{\MakeTextUppercase}{}{}
\makeatother
\fi

\usepackage[ruled]{algorithm2e}

\SetAlFnt{\small}
\SetAlCapFnt{\small}
\SetAlCapNameFnt{\small}
\SetAlCapHSkip{0pt}
\IncMargin{-\parindent}

\usepackage{amssymb, amsmath, amsthm}
\usepackage{epsfig,graphicx}

\usepackage{color-edits}

\addauthor{ar}{red}
\addauthor{mb}{blue}
\addauthor{nn}{green}

\usepackage{tikz, ifthen,etoolbox,color}
\usepackage{calc}
\usepackage{float}
\def\PAPER{1}
\ifFULL
\def\SCALEA{0.05}
\def\SCALEB{0.055}
\def\SCALEC{0.05}
\def\MINIPAGEWIDTH{0.49}
\else
\def\SCALEA{0.03}
\def\SCALEB{0.035}
\def\SCALEC{0.03}
\def\MINIPAGEWIDTH{0.46}
\fi

\def\FIGURECAPTION{\footnotesize}

\ifFULL
\newtheorem{theorem}{Theorem}[section]
\newtheorem{lemma}[theorem]{Lemma}
\newtheorem{corollary}[theorem]{Corollary}

\fi

\newtheorem{claim}[theorem]{Claim}

\theoremstyle{definition}
\ifFULL
\newtheorem{definition}[theorem]{Definition}
\newtheorem{example}[theorem]{Example}
\fi
\newtheorem{remark}[theorem]{Remark}
\newtheorem{oq}[theorem]{Open Question}

\newcommand {\ignore} [1] {}

\newcommand{\srev}{\textsc{SRev}}

\newcommand{\brev}{\textsc{BRev}}
\newcommand{\rev}{\textsc{Rev}}
\newcommand{\drev}{\textsc{DRev}}

\newcommand{\smdrev}{\textsc{SMRev}}
\newcommand{\DD}{\mathcal{D}}
\newcommand{\FF}{\mathcal{F}}

\newcommand {\ERr} [1] {{{\textsc {ER}}(#1)}}

\begin{document}
\title{Optimal Deterministic Mechanisms for an Additive Buyer}
\ifFULL
\author{Moshe Babaioff\thanks{Microsoft Research, \texttt{moshe@microsoft.com}}	
\and 	Noam Nisan\thanks{Hebrew University and Microsoft Research, \texttt{noam@cs.huji.ac.il}} 	
\and 	Aviad Rubinstein\thanks{Harvard University, \texttt{aviad@seas.harvard.edu}}}
\else
\author{Submission 89}
\fi		
		
\ifFULL
\maketitle
\fi
\begin{abstract}
	
	We study revenue maximization by {\em deterministic} mechanisms for the simplest case for which Myerson's characterization does not hold: a single seller selling two items, with independently distributed values, to a single additive buyer.  
	We prove that optimal mechanisms are submodular and hence monotone. 
	Furthermore, we show that in the IID case, optimal mechanisms are symmetric.
	Our characterizations are surprisingly non-trivial, and we show that they fail to extend
	in several natural ways, e.g.~for correlated distributions or more than two items. 
	In particular, this shows that the optimality of symmetric mechanisms does \emph{not} follow from the symmetry of the IID distribution.

\end{abstract}

\ifFULL
\else
\maketitle
\renewcommand\LARGE{\small}		
\fi		
		

\section{Introduction}
This paper deals with the problem of revenue maximization by a monopolist seller in the simple scenario when a single buyer has an additive valuation over a set of items.
Our main focus is on \emph{deterministic mechanisms}, 
and we start with the simplest case when there are just two items for sale. 
In that case, the buyer's value for the first item is $v_1\geq 0$, for the second item is $v_2\geq 0$, and for the bundle of both is $v_1+v_2$, where the seller
only knows that $v_1$ is sampled from distribution $\DD_1$ and $v_2$ is sampled from distribution $\DD_2$, and that these values
are independently sampled.  Our seller's goal is to maximize his expected revenue when the buyer maximizes his quasi-linear utility.

This simplest of scenarios has received much attention lately as it is one of the most fundamental examples where Myerson's 
characterization of optimal auctions 
(for a single item) \cite{Myerson81} ceases to hold, and indeed optimal 
auctions for two items may become complex \cite{MCAFEE19885,AlaeiFHHM12, 
	ManelliV07, HartN12, HR15-nonmon, DDT15-infinite_menu}.  
In particular, different examples are known where the optimal auction sells each of the two items separately,
sells both items as a bundle, gives a ``discount price'' for the bundle \cite{HartN12}, is randomized (i.e. uses
lotteries) \cite{ManelliV07}, requires an infinite number of possible randomized outcomes \cite{DDT15-infinite_menu}, or is
non-monotone (i.e. the revenue it extracts may decrease when the buyer's valuations increase.) \cite{HR15-nonmon}.

Generally speaking we do not understand the structure of optimal auctions -- neither randomized ones
nor deterministic ones -- even
in this simple scenario, and in fact
it is known that for the case of $n$ items (still with a single additive buyer) determining the optimal randomized auction
is $\#P$-hard \cite{DaskalakisDT14} 
as is determining the optimal deterministic auction \cite{CMPY17-complexity-drev}.\footnote{\citet{ConitzerS04} have proven NP hardness of optimal deterministic auction for a much more general setting.}
On the other hand, several recent results show that simple auctions (such as selling the items separately or as
a single bundle) provide good approximations to the optimal revenue for two items \cite{HartN12}, multiple items
\cite{LY13, BILW14}, as well as further generalizations in scenario such as multiple buyers \cite{Yao15}, or combinatorial valuations \cite{RW15-subadditive}, or both \cite{CM16-many, CZ17-many_XOS, Yao17-monotone} (but {\em not} when the item values are correlated \cite{HartN13} --- except for special cases~\cite{BateniDHS15}).

In this paper we prove structural characterizations of the optimal {\em deterministic}
mechanisms for a monopolist seller that is selling {\em two items} to a single\footnote{Note that results for a single buyer also hold when there are many buyers but there are no supply constraints (digital goods).}
additive buyer with item values \emph{independently distributed}.\footnote{A deterministic mechanism for such a problem simply presents prices for each of the items, and for the pair.
}
We also show that these characterizations fail to generalize in any conceivable way: 
neither to 
more than two items, nor to items with correlated distributions, nor to randomized optimal mechanisms.  

\begin{figure}[!t] 
	\caption{Submodular vs Supermodular Menus}\label{two-types}
	\vspace{0.3cm}


\hspace{-0.4cm}
	\begin{minipage}[t]{\MINIPAGEWIDTH\columnwidth}
		\begin{center}
			\ifdefined\PAPER
\else

\documentclass[10pt]{article}
\usepackage{tikz, ifthen,etoolbox,color}
\usepackage{calc}
\usetikzlibrary{arrows.meta}
\newcommand{\ignore}[1]{}%

\begin{document}
\pagestyle{empty}
\fi
\ifdefined\SCALEC
\else
\def\SCALEC{0.07}
\fi

\ifdefined\CHUPCHIK
\else
\def\CHUPCHIK{6}
\def\cwidth{1}
\fi

\begin{tikzpicture} [scale = \SCALEC]
\tikzstyle{every node}=[font=\LARGE]
\usetikzlibrary{calc}

\def\a{15} 
\def\b{45} 
\def\c{80} 

\def\width{3} 
\def\shift{1.5} 

\draw  (0,0) -- (0,100) -- (100,100)-- (100,0) -- (0,0) [line width = \width];

\draw  (0,\b) -- (\a, \b) -- (\a, 0) [line width = \width];
\draw  (\c-\b,100) -- (\c-\b,\c-\a) -- (100,\c-\a)  [line width = \width];
\draw  (\a,\b) -- (\c-\b,\c-\a)  [line width = \width];

\draw  (-\CHUPCHIK,\b) -- (0, \b) [line width = \cwidth]; 
\draw  (\a, 0) -- (\a, -\CHUPCHIK) [line width = \cwidth]; 
\draw  (\c-\b,100+\CHUPCHIK) -- (\c-\b,100) [line width = \cwidth]; 
\draw  (100,\c-\a) -- (100+\CHUPCHIK,\c-\a)  [line width = \cwidth]; 

\draw (\a, -4 -\CHUPCHIK) node {$a$};
\draw (-4-\CHUPCHIK, \b) node {$b$};
\draw (\c-\b, 100 + 4+\CHUPCHIK) node {$c-b$};
\draw (110 + 4 +\CHUPCHIK, \c-\a) node {$c-a$};

\draw  (0,0) -- (0,115) [line width = 0.5*\width, dashed, ->]; 
\draw (-8, 110) node {{\color{blue}$v_2$}};
\draw  (0,0) -- (115,0) [line width = 0.5*\width, dashed, ->]; 
\draw (110,-8) node {{\color{blue}$v_1$}};

\draw (50+0.5*\a, 0.5*\c - 0.5*\a) node {$a$};
\draw (0.5*\c - 0.5*\b, 50+0.5*\b) node {$b$};
\draw (50 + 0.5*\c - 0.5*\b, 50+0.5*\c - 0.5*\a) node {$c$};

\end{tikzpicture}

\ifdefined\PAPER
\else
\end{document}
\fi
			
			{\FIGURECAPTION Supermodular menu $(a,b,c)$, where $c\geq a+b$. }
		\end{center}
	\end{minipage}
	\begin{minipage}[t]{\MINIPAGEWIDTH\columnwidth}
		\begin{center}
			\ifdefined\PAPER
\else

\documentclass[10pt]{article}
\usepackage{tikz, ifthen,etoolbox,color}
\usepackage{calc}
\usetikzlibrary{arrows.meta}
\newcommand{\ignore}[1]{}%

\begin{document}
\pagestyle{empty}

\fi

\ifdefined\SCALEA
\else
\def\SCALEA{0.07}
\fi
\begin{tikzpicture} [scale = \SCALEA]
\tikzstyle{every node}=[font=\LARGE]
\usetikzlibrary{calc}

\def\a{27} 
\def\b{70} 
\def\c{85} 

\def\width{3} 
\def\shift{1.5} 

\ifdefined\CHUPCHIK
\else
\def\CHUPCHIK{6}
\def\cwidth{1}
\fi

\draw  (0,0) -- (0,100) -- (100,100)-- (100,0) -- (0,0) [line width = \width];

\draw  (0,\b) -- (\c-\b,\b) -- (\c-\b,100)  [line width = \width];
\draw  (\a,0) -- (\a,\c-\a) -- (100,\c-\a)  [line width = \width];
\draw  (\a,\c-\a) -- (\c-\b,\b)  [line width = \width];

\draw  (-\CHUPCHIK,\b) -- (0, \b) [line width = \cwidth]; 
\draw  (\a, 0) -- (\a, -\CHUPCHIK) [line width = \cwidth]; 
\draw  (\c-\b,100+\CHUPCHIK) -- (\c-\b,100) [line width = \cwidth]; 
\draw  (100,\c-\a) -- (100+\CHUPCHIK,\c-\a)  [line width = \cwidth]; 

\draw (0.5*\a+50, 0.5*\c-0.5*\a) node {$a$};
\draw (0.5*\c-0.5*\b, 0.5*\b+50) node {$b$};
\draw (0.5*\c-0.5*\b+50, 0.5*\c-0.5*\a+50) node {$c$};

\draw (\a, -7-\CHUPCHIK) node {$a$};
\draw (-7-\CHUPCHIK, \b) node {$b$};
\draw (\c-\b, 107+\CHUPCHIK) node {$c-b$};
\draw (101+\CHUPCHIK,\c-\a) node [right] {$c-a$};

\end{tikzpicture}
\ifdefined\PAPER
\else
\end{document}
\fi
			
			{\FIGURECAPTION Submodular menu $(a,b,c)$, where $c\leq a+b$.}
		\end{center}
	\end{minipage}\hfill{}%

\vspace{0.5cm}
{\FIGURECAPTION 
	Given a menu, the incentive constraints of the buyer imply a partition of the space of valuations to four regions.
	The allocation is empty in the South-West region and the revenue is 0. The first item is allocated in the South-East region, the second in the North-West region, and the pair in the North-East region. In each of these regions we mark the revenue for any valuation in that region. We omit the allocation, and payment when there is no sale, in all figures, as they are the same as here. Note that the borders between any two regions are places of indifference, and as ties are broken towards higher payments, each belongs to the neighboring region with the higher payment. The location of the intersection of each of these borders with the axes is also marked (e.g. at $(a,0)$ the buyer is indifferent between buying nothing and buying the first item.)
Note that while our results are not limited to bounded valuations, our figures are drawn as if they are upper bounded, for convenience of illustration. 
Finally, to avoid clutter, we explicitly draw the axes only for the figure corresponding to the first (supermodular) menu, and eliminate them from all  other figures.}
\end{figure}
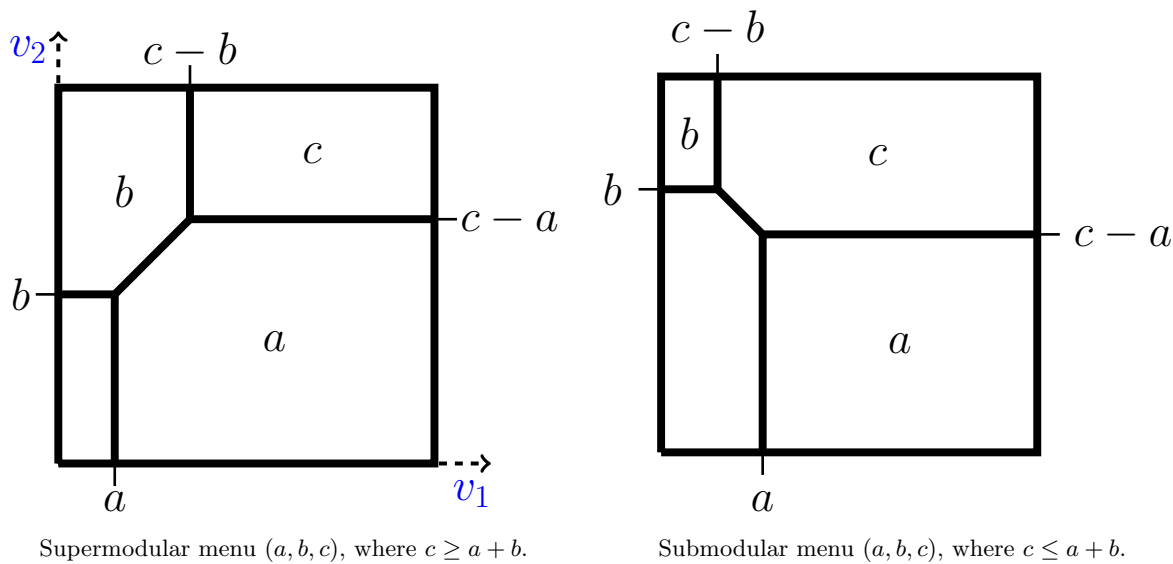

A general deterministic 
auction for two items charges prices, $a$ and $b$ respectively, for each of the two items, as well as a third price $c$
for the bundle of both items.  
There are basically two ``forms'' of such auctions: the first is the
``submodular case'' where $c \le a+b$ and the second is the ``supermodular case'' where $c \ge a+b$.\footnote{When $c=a+b$ the menu is additive - it is both submodular and supermodular. Additive pricing corresponds to selling the item separately.}
Given a menu, the incentive constraints of the buyer imply a partition of the space of valuations to regions, in each the buyer buys the same bundle.
The different shapes of the partition that these two types of auctions
induce on the buyer's two-item value space are depicted in Figure \ref{two-types}.  
Our first result states that the seller never loses by presenting the buyer with an auction 
of the submodular type.


\begin{theorem} \label{thm:intro-main}	
	For every two distributions $\DD_1$ and $\DD_2$, and a single additive buyer with two items' valuations sampled from $\DD_1 \times \DD_2$, 
	the revenue of any deterministic mechanism can be obtained by a deterministic mechanism that is \emph{submodular}.
\end{theorem}
This means that the seller will never gain by posting a price for the pair that is higher than the sum of prices for the two items. 
This is highly desirable property as it implies that in such a mechanism 
a buyer cannot gain by using ``false names'' \cite{YokooSM04} and buying each of the two items separately under a different identity (e.g. when the seller 
of a digital good faces multiple ex-ante identical buyers, and she posts the same prices to every buyer.) One practical motivation for false-name proof mechanisms is that preventing the usage of false names requires verification of identities, which might be very costly. Another motivation is that for some mechanisms, allowing the buyer to select multiple menu options may result in exponential savings in representation size\footnote{For example, item pricing can be represented with linear menu in the case multiple menu entries can be picked, but require exponential menu when the buyer can only pick a single entry from the menu}; however such a representation only makes sense for false-name-proof mechanisms. (See also Remark~\ref{rem:menu-size} for further discussion on ``menu-size'' vs ``additive-menu-size''.) 

An additional benefit of such a mechanism is that it satisfies revenue monotonicity - it was shown by \cite{HR15-nonmon} that submodular deterministic auctions are revenue monotone\footnote{For the case of two items it is actually easy to see that any submodular deterministic auction is revenue monotone: increasing values cannot cause the bidder to switch from one item to the other, only from purchasing one item to buying the pair, and the price of the pair is at least the price of any item.}
(i.e. the revenue that they extract from a player with item values $(v_1,v_2)$ cannot be lower than the revenue that they extract from a player
with values $(v_1',v_2')$ with $v_1' \le v_1$ and $v_2' \le v_2$). From the theorem and \cite{HR15-nonmon} we immediately get the next corollary.

\begin{corollary} \label{cor:intro-rev-mon}
	For every two distributions $\DD_1$ and $\DD_2$, and a single additive buyer with two items' valuations sampled from $\DD_1 \times \DD_2$,
	the revenue of any deterministic mechanism can be obtained by a deterministic mechanism that is \emph{revenue-monotone}.
\end{corollary} 

Using Corollary~\ref{cor:intro-rev-mon}
we are also able to determine exactly the maximum possible gap between the optimal revenue by deterministic auctions
and the optimal revenue from selling the items separately.  If we denote the optimal revenue of deterministic auctions by $\drev(\DD_1 \times \DD_2)$ 
and the optimal revenue from selling the two items separately (each at its optimal Myerson price) by $\srev(\DD_1 \times \DD_2)$ then
we get:

\begin{theorem}\label{thm:intro-1.278}
	For every two distributions $\DD_1$ and $\DD_2$ and a single additive buyer with two items' valuations sampled from $\DD_1 \times \DD_2$, 
	it holds that $\drev(\DD_1 \times \DD_2) \le w \cdot \srev(\DD_1 \times \DD_2)$, where
	$w \approx 1.278$ is the solution of $(w-1) \cdot e^w=1$. 
\end{theorem}

This bound is tight as \cite{HartN12} shows that for an ``equal revenue distribution\footnote{A distribution for which any price in the support has the same revenue. For a formal definition see Section \ref{sec:main}.}'' $ER$ it holds that
$\drev(ER \times ER) = w \cdot \srev(ER \times ER)$.  It is still open whether the gap between $\srev$ and $\rev$ (the revenue of the optimal \emph{randomized} mechanism) is larger or not.

We then note several ways in which Theorem \ref{thm:intro-main}, which holds 
for deterministic auctions for an additive buyer with two independent items, 
can {\em not} be generalized:

\begin{enumerate}
	\item {\bf To more than two items}: We exhibit an example where the optimal deterministic auction for three independent items fails to be submodular.\footnote{For a set of items $M$, a pricing function $p:2^M\rightarrow \mathbb{R}$ is \emph{submodular} if for any $S,T\subseteq M$ it holds that $p(S)+p(T)\geq p(S\cap T) + p(S\cup T)$.}  
	We leave as an open problem whether optimal deterministic auctions for more than two items must at least be 
	subadditive.\footnote{For a set of items $M$, a pricing function $p:2^M\rightarrow \mathbb{R}$ is \emph{subadditive} if for any $S,T\subseteq M$ it holds that $p(S)+p(T)\geq p(S\cup T)$.}  
	\item {\bf To correlated item values}: 
	We show that the optimal deterministic auction for correlated distributions on the
	two items may fail to be submodular (or monotone).  
	Nevertheless, the gap between the two is bounded: we show that the maximum possible gap between the revenues of general
	deterministic versus submodular deterministic%
\footnote{Our result is actually a little stronger: we show that the gap is at most $3/2$ even if the submodular auction is restricted to either selling each item separately or offering only the grand bundle.}
 auctions in the case of two items with correlated distributions is $3/2$, and this is essentially tight.	
	\item {\bf To randomized auctions}: while it is not clear what is the exact analog of submodularity for a randomized auction,
	it is known \cite{HR15-nonmon} that the optimal \emph{randomized} auction for two independently distributed items need not be revenue monotone, 
	in contrast to the revenue monotonicity result that we present above for \emph{deterministic} auction (Corollary \ref{cor:intro-rev-mon}).
	We also note that while submodular auctions for two items are ``false name proof'', optimal randomized auctions need not be so. 
\end{enumerate}

Our second structural result concerns the question of symmetry.  Suppose that the two items are symmetric, 
i.e. sampled IID from the distribution $\DD_1=\DD_2=\FF$.
Does this also imply that there is an optimal deterministic auction that is symmetric (both items have the same price)?  For general (not necessarily deterministic) auctions,
symmetry may be obtained without loss of generality by averaging over all possible permutations of the items, but this 
procedure results in a randomized auction even when starting with a deterministic one.  It turns out that the answer is
still positive, but this is rather delicate.

\begin{theorem}\label{thm:intro-symmetric}
	For any distribution $\FF$, and a single additive buyer with two items' valuations sampled from $\FF\times \FF$, 
	the revenue of any deterministic mechanism can be obtained by a deterministic mechanism that is \emph{symmetric}.
\end{theorem}

Perhaps surprisingly, we show that this theorem can \emph{not} be generalized in each of these natural ways:

\begin{enumerate}
	\item {\bf To correlated symmetric item values}: We show that there exists a correlated symmetric joint distribution on $(v_1,v_2)$ 
	such that the revenue of the best symmetric deterministic auction is strictly less than that of the optimal
	deterministic auction.  
	\item {\bf To more than two IID items}: We exhibit an example of a distribution $\FF$ where for three items whose values are distributed,
	independently, according to $\FF$, the optimal symmetric deterministic auction extracts strictly less revenue than the 	optimal deterministic auction. 
\end{enumerate}
In particular, the above examples show that the symmetry of the joint distribution 
that is implied by the IID assumption is \emph{not} sufficient to derive that the optimal pricing is symmetric.

It should be noted that in contrast to the failure of the positive result for \emph{deterministic} 2-item case to generalize to joint correlated distributions and to more items,  
for the case of \emph{randomized} auctions and any number of items, it is trivially true that 
for any symmetric joint distribution over $n$ items, 
there exists an optimal randomized auction that is symmetric (by going over all item permutations.) 
This is true when item values are samples IID, and it is also true when the items' values are correlated in a symmetric way.

\subsection{Additional related work}
There are countless papers dedicated to the topics we touch on in this work, including tradeoffs between simple and complex mechanisms, structure of optimal mechanisms, false-name proofness, etc. We shall briefly mention a few below.
\citet{ThirumulanathanSN2016} study revenue maximization for the special case uniforms distributions on intervals of the real line. \citet{ConitzerS04} prove computational intractability albeit for the much more general setting of correlated valuations. \cite{Thanassoulis04, ChawlaHK07, BriestCKW15} study questions related to ours, but for buyers with unit-demand valuations (whereas we focus on additive valuations). 
	


\section{Preliminaries}
A monopolist seller sets up a mechanism to maximize revenue when facing a single buyer. 
The buyer has an additive valuation over $n$ items, that is, if his value for item $i$ is $v_i\geq 0$, 
his value for the set of items $S$ is $v(S) =\sum_{i\in S} v_i$.  
The values of the items are drawn from a distribution $\DD$ which is known to the seller. 
We mostly focus on the special cases where $n=2$ and the value $v_i$ for each item $i$ is drawn independently from a distribution $\DD_i$, and $\DD=\DD_1\times \DD_2\times\ldots \times\DD_n$.

We study the revenue that can be achieved by a monopolist seller facing such a buyer. 
We consider deterministic  mechanisms that are \emph{truthful} and ex-post Individually Rational (IR), which for a single buyer are simply menus that present a price for each bundle, and the buyer picks a bundle which maximizes his quasi-linear utility (the difference between the value for the bundle and the payment), breaking ties in favor of higher payments.


\subsubsection*{Notation}
For any joint distribution $\DD$ of valuations, we use the following
notation, mostly due to \cite{HartN12,BILW14}, to denote the
optimum revenue for each class of mechanisms:
\begin{itemize}
\item $\rev\left(\DD\right)$ - the supremum revenue among all truthful
mechanisms;
\item $\drev\left(\DD\right)$ - the supremum revenue among all truthful
{\em deterministic} mechanisms;
\item $\brev\left(\DD\right)$ - the supremum revenue obtainable by pricing
only the grand bundle;\footnote{Pricing every non-empty bundle at the same price - WLOG the buyer will either buy all the items, or nothing.} and
\item $\srev\left(\DD\right)$ - the supremum revenue obtainable by pricing each item separately;\footnote{Every bundle is priced at the sum of prices of the items in the bundle.}
\item $\smdrev\left(\DD\right)$ - the supremum revenue among all truthful deterministic mechanisms with submodular pricing.
\end{itemize}
When $\DD$ is clear from the context, we simply write $\rev,\drev,$
etc.

We note that the supremum might not be obtained by any mechanism, even when a single item is for sale 
to a single buyer, and even if the optimal revenue is bounded from above. 
To see this, consider the distribution with support over all non-negative real numbers and for which $Pr[v\geq p]=1/(1+p)$, for such a distribution the revenue with price $p$ is $p/(1+p)$ which grows to $1$ as $p$ goes to infinity, yet no price obtains revenue of $1$. 
On the other hand, for any distribution with a finite support the supremum is clearly obtained. 
When the supremum is obtained, each of our results that mechanisms with some property $X$ (e.g. submodular) can get the same revenue as any mechanisms, implies that there is an optimal mechanism with property $X$.



A deterministic menu for 2 items can be represented by 3 prices: two prices for each of the two items, and another price for the pair. We use triplets as $(p_1,p_2,p_{1,2})$ to denote such a menu, with $p_1$ and $p_2$ being the prices for the first and second item, respectively, and $p_{1,2}$ being the price for the pair, with all prices being non-negative.
Note also that without loss of generality, the seller can restrict her attention to menus with $p_{1,2}\geq \max\{p_1,p_2\}$ as decreasing the price of an item to the price of the pair never decreases the revenue. 
We use $\rev_\DD(x,y,z)$ to denote the expected revenue from menu $(x,y,z)$ for the distribution $\DD$. When the distribution is clear from the context, we use the simpler notation $\rev(x,y,z)$.

\subsubsection*{Revenue Monotonicity}
A menu is {\em revenue monotone} if the revenue obtained by the menu does not decrease when the value of every item weakly increases. That is,
a revenue monotone menu satisfies the condition that the revenue from a player with item values $(v_1,v_2)$ cannot be lower than the revenue extracted from a player
with values $(v_1',v_2')$ with $v_1' \le v_1$ and $v_2' \le v_2$.

\section{Deterministic Pricing for Two Independent Items}\label{sec:main}
We prove that for two independent items and an additive buyer, a deterministic seller does not lose revenue by restricting herself to mechanisms that are \emph{submodular}.
We do so by showing that the revenue of any supermodular menu is dominated by the revenue of one out of two additive menus that are derived from the supermodular menu.  
\begin{theorem}[Theorem \ref{thm:intro-main} restated]\label{thm:submodular}
	For every two distributions $\DD_1$ and $\DD_2$, and a single additive buyer with two items' valuations sampled from $\DD_1 \times \DD_2$, the revenue of any deterministic mechanism can be obtained by a deterministic mechanism that is \emph{submodular}. Thus, 
	$\smdrev(\DD_1 \times \DD_2) = \drev(\DD_1 \times \DD_2)$,	 
\end{theorem}

\begin{proof}
Given a 
strictly 
supermodular menu $(a,b,c)$ (selling item $1$, item $2$, and the grand bundle for prices $a,b,c$, respectively), assume wlog that 
$a \leq b$. By strict supermodularity, $c>a+b\geq b$.
We
argue that at least one of the following additive menus achieves at least as much revenue: selling the items separately for prices $a,b$, respectively (menu $(a,b,a+b)$), and selling separately for prices $c-b,c$ (menu ($c-b,b,c$)).\footnote{It may not be immediately intuitive why charging $c-b$ for the first item is a good idea. However, we expect to have probability mass at $v_1 \approx c-b$ because this is where, when facing the original supermodular menu, the buyer switches from buying item $2$ to the grand bundle.}

The first step of the proof is to consider the partition of the valuation space induced by the incentive compatibility constraints of the three menus, and to compute the payments in each region (see Figure~\ref{fig:submodular-proof} for details).
We now split into cases depending on the ratio between $\Pr\big[v_1 \in [a, c-b) \big]$ and $\Pr\big[v_1 \geq c-b\big]$. (Notice that this step only makes sense for independent valuations.) 
We show that when 
\vspace{-0.9cm}

\begin{figure}
	\caption{Submodular menus are optimal}\label{fig:submodular-proof}
	\vspace{1.5cm}
	\hspace*{\fill}
	\begin{minipage}[t]{0.49\columnwidth}
		\begin{center}
			\ifdefined\PAPER
\else

\documentclass[10pt]{article}
\usepackage{tikz, ifthen,etoolbox,color}
\usepackage{calc}
\usetikzlibrary{arrows.meta}
\newcommand{\ignore}[1]{}%

\begin{document}
\pagestyle{empty}
\fi
\ifdefined\SCALEC
\else
\def\SCALEC{0.07}
\fi
\begin{tikzpicture} [scale = \SCALEC]
\tikzstyle{every node}=[font=\LARGE]
\usetikzlibrary{calc}

\def\a{15} 
\def\b{45} 
\def\c{80} 

\def\width{3} 
\def\shift{1.5} 

\ifdefined\CHUPCHIK
\else
\def\CHUPCHIK{6}
\def\cwidth{1}
\fi

\draw  (0,0) -- (0,100) -- (100,100)-- (100,0) -- (0,0) [line width = \width];

\draw  (0,\b) -- (\a, \b) -- (\a, 0)  [line width = \width];
\draw  (\c-\b,100) -- (\c-\b,\c-\a) -- (100,\c-\a)  [line width = \width];
\draw  (\a,\b) -- (\c-\b,\c-\a)  [line width = \width];

\draw  (-\CHUPCHIK,\b) -- (0, \b) [line width = \cwidth]; 
\draw  (\a, 0) -- (\a, -\CHUPCHIK) [line width = \cwidth]; 
\draw  (\c-\b,100+\CHUPCHIK) -- (\c-\b,100) [line width = \cwidth]; 
\draw  (100,\c-\a) -- (100+\CHUPCHIK,\c-\a)  [line width = \cwidth]; 

\draw (\a, -7 -\CHUPCHIK) node {$a$};
\draw (-7-\CHUPCHIK, \b) node {$b$};
\draw (\c-\b, 107+\CHUPCHIK) node {$c-b$};
\draw (116 +\CHUPCHIK, \c-\a) node {$c-a$};

\draw (50+0.5*\a, 0.5*\c - 0.5*\a) node {$a$};
\draw (0.5*\c - 0.5*\b, 50+0.5*\b) node {$b$};
\draw (50 + 0.5*\c - 0.5*\b, 50+0.5*\c - 0.5*\a) node {$c$};

\end{tikzpicture}

\ifdefined\PAPER
\else
\end{document}
\fi
			
			{\FIGURECAPTION Payments to seller with strictly supermodular menu $(a,b,c)$, satisfying $c>a+b$.}
		\end{center}
	\end{minipage}\hfill{}%
	\begin{minipage}[t]{0.49\columnwidth}
		\begin{center}
			\ifdefined\PAPER
\else

\documentclass[10pt]{article}
\usepackage{tikz, ifthen,etoolbox,color}
\usepackage{calc}
\usetikzlibrary{arrows.meta}
\newcommand{\ignore}[1]{}%

\begin{document}
\pagestyle{empty}
\fi
\ifdefined\SCALEC
\else
\def\SCALEC{0.07}
\fi
\begin{tikzpicture} [scale = \SCALEC]
\tikzstyle{every node}=[font=\LARGE]
\usetikzlibrary{calc}

\def\a{15} 
\def\b{45} 
\def\c{80} 

\def\width{3} 
\def\shift{0} 

\ifdefined\CHUPCHIK
\else
\def\CHUPCHIK{6}
\def\cwidth{1}
\fi

\draw  (0,0) -- (0,100) -- (100,100)-- (100,0) -- (0,0) [line width = \width];

\draw  (0,\b+\shift) -- (100,\b+\shift)  [line width = \width, color = brown];
\draw  (\a+\shift,0) -- (\a+\shift,100)  [line width = \width, color = brown];

\draw  (-\CHUPCHIK,\b) -- (0, \b) [line width = \cwidth]; 
\draw  (\a, 0) -- (\a, -\CHUPCHIK) [line width = \cwidth]; 


\draw (\a, -7-\CHUPCHIK) node {$a$};
\draw (-7-\CHUPCHIK, \b) node {$b$};

\draw (50+0.5*\a, 0.5*\b) node {\color{brown}$a$};
\draw (0.5*\a, 50+0.5*\b) node {\color{brown}$b$};
\draw (50 + 0.5*\a, 50+0.5*\b) node {\color{brown}$a+b$};

\end{tikzpicture}

\ifdefined\PAPER
\else
\end{document}
\fi
			
			{\FIGURECAPTION Payments to seller with the additive menu {\color{brown}$(a,b,a+b)$}.}
		\end{center}
	\end{minipage}\hspace*{\fill}
	
	\vspace{1.5cm}
	\hspace*{\fill}
	\begin{minipage}[t]{0.49\columnwidth}
		\begin{center}
			\ifdefined\PAPER
\else

\documentclass[10pt]{article}
\usepackage{tikz, ifthen,etoolbox,color}
\usepackage{calc}
\usetikzlibrary{arrows.meta}
\newcommand{\ignore}[1]{}%

\begin{document}
\pagestyle{empty}
\fi
\ifdefined\SCALEC
\else
\def\SCALEC{0.07}
\fi
\begin{tikzpicture} [scale = \SCALEC]
\tikzstyle{every node}=[font=\LARGE]
\usetikzlibrary{calc}

\def\a{15} 
\def\b{45} 
\def\c{80} 

\def\width{3} 
\def\shift{1.5} 

\ifdefined\CHUPCHIK
\else
\def\CHUPCHIK{6}
\def\cwidth{1}
\fi

\draw  (0,0) -- (0,100) -- (100,100)-- (100,0) -- (0,0) [line width = \width];

\draw  (0,\b-\shift) -- (100,\b-\shift)  [line width = \width, color = blue];
\draw  (\c-\b-\shift,0) -- (\c-\b-\shift,100)  [line width = \width, color = blue];


\draw  (-\CHUPCHIK,\b) -- (0, \b) [line width = \cwidth]; 
\draw  (\a, 0) -- (\a, -\CHUPCHIK) [line width = \cwidth]; 

\draw (\a, -7 -\CHUPCHIK) node {$a$};
\draw (-7-\CHUPCHIK, \b) node {$b$};

\draw (50+0.5*\c-0.5*\b, 0.5*\b) node {\color{blue}$c-b$};
\draw (0.5*\c-0.5*\b, 50+0.5*\b) node {\color{blue}$b$};
\draw (50 + 0.5*\c-0.5*\b, 50+0.5*\b) node {\color{blue}$c$};

\end{tikzpicture}

\ifdefined\PAPER
\else
\end{document}
\fi
			
			{\FIGURECAPTION Payments to seller with the additive menu {\color{blue}$(c-b,b,c)$}.}
		\end{center}
	\end{minipage}\hfill{}%
	\begin{minipage}[t]{0.49\columnwidth}
		\begin{center}
			\ifdefined\PAPER
\else

\documentclass[10pt]{article}
\usepackage{tikz, ifthen,etoolbox,color}
\usepackage{calc}
\usetikzlibrary{arrows.meta}
\newcommand{\ignore}[1]{}%

\begin{document}
\pagestyle{empty}
\fi
\ifdefined\SCALEC
\else
\def\SCALEC{0.07}
\fi
\begin{tikzpicture} [scale = \SCALEC]
\tikzstyle{every node}=[font=\LARGE]
\usetikzlibrary{calc}

\def\a{15} 
\def\b{45} 
\def\c{80} 

\def\width{3} 
\def\shift{1.5} 

\ifdefined\CHUPCHIK
\else
\def\CHUPCHIK{6}
\def\cwidth{1}
\fi

\draw  (0,0) -- (0,100) -- (100,100)-- (100,0) -- (0,0) [line width = \width];

\draw  (0,\b) -- (\a, \b) -- (\a, 0)  [line width = \width];
\draw  (\c-\b,100) -- (\c-\b,\c-\a) -- (100,\c-\a)  [line width = \width];
\draw  (\a,\b) -- (\c-\b,\c-\a)  [line width = \width];

\draw  (-\CHUPCHIK,\b) -- (0, \b) [line width = \cwidth]; 
\draw  (\a, 0) -- (\a, -\CHUPCHIK) [line width = \cwidth]; 
\draw  (\c-\b,100+\CHUPCHIK) -- (\c-\b,100) [line width = \cwidth]; 
\draw  (100,\c-\a) -- (100+\CHUPCHIK,\c-\a)  [line width = \cwidth]; 

\draw  (0,\b+\shift) -- (100,\b+\shift)  [line width = \width, color = brown];
\draw  (\a+\shift,0) -- (\a+\shift,100)  [line width = \width, color = brown];

\draw  (0,\b-\shift) -- (100,\b-\shift)  [line width = \width, color = blue];
\draw  (\c-\b-\shift,0) -- (\c-\b-\shift,100)  [line width = \width, color = blue];


\draw (\a, -7-\CHUPCHIK) node {$a$};
\draw (-7-\CHUPCHIK, \b) node {$b$};
\draw (\c-\b, 107+\CHUPCHIK) node {$c-b$};
\draw (114+\CHUPCHIK, \c-\a) node {$c-a$};

\draw (50+0.5*\c-0.5*\b, 0.5*\b) node {\color{blue}$c-(a+b)$};
\draw (0.5*\a+0.5*\c-0.5*\b, 0.5*\b) node {\color{blue}$-a$};

\draw (0.5*\a+0.5*\c-0.5*\b, 50+0.5*\b) node {\color{brown}$a$};
\draw (50 + 0.5*\a, 0.5*\c-0.5*\a+0.5*\b) node {\color{brown}$b$};
\draw (0.5*\a+0.5*\c-0.4*\b, 0.5*\c-0.5*\a+0.45*\b) node {\color{brown}$b$};
\draw (50 + 0.5*\c-0.5*\b, 50 + 0.5*\c-0.5*\a) node {\color{brown}$(a+b)-c$};

\end{tikzpicture}

\ifdefined\PAPER
\else
\end{document}
\fi
			{\FIGURECAPTION 
	In each region, we compute the difference in revenues of two menus.
	
Above $v_2 = b$: {\color{brown}$(a,b,a+b)$} vs. $(a,b,c)$
			({\color{blue}$(c-b,b,c)$} dominates $(a,b,c)$).

			Below $v_2 = b$:  {\color{blue}$(c-b,b,c)$} vs. $(a,b,c)$ ({\color{brown}$(a,b,a+b)$} and $(a,b,c)$ are the same).}
			
			
		\end{center}
	\end{minipage}\hspace*{\fill}
	
\end{figure}
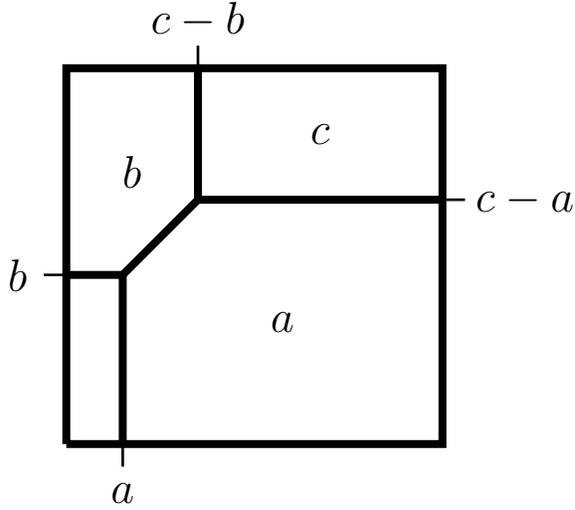


\begin{gather}\label{eq:Submodular-case1} \Pr\big[v_1 \in [a, c-b) \big] \cdot a \geq \Pr\big[v_1 \geq c-b\big] \cdot \big(c-(a+b)\big), \end{gather}
the menu $(a,b,a+b)$ obtains at least as much revenue as menu $(a,b,c)$,
while when this is violated, the  
menu $(c-b,b,c)$ obtains at least as much revenue as menu $(a,b,c)$.

\subsubsection*{Detailed explanations}
Below we explain why in each case the respective submodular menu is at least as good as supermodular menu $(a,b,c)$. Refer also to Figure~\ref{fig:submodular-proof}.

First, we argue that when Inequality~\eqref{eq:Submodular-case1} holds, menu $(a,b,a+b)$ obtains at least as much revenue as menu $(a,b,c)$.
This is so as both menus obtain the same revenue for any valuations with $v_2\leq b$,  
and for any $v_2>b$, due to independence, the contribution to the expected gain of the event $v_1 \in [a, c-b)$ is at least as large as the contributed expected loss of the event  $v_1 \geq c-b$
(note that for $v_2$ such that $b\leq v_2\leq c-a$, the menu $(a,b,a+b)$ obtains at least as much revenue as the menu $(a,b,c)$, so it is enough to look at $v_2>c-a$ and use independence).

Finally, we argue that when Inequality~\eqref{eq:Submodular-case1} fails to hold, 
menu $(c-b,b,c)$ obtains at least as much revenue as menu $(a,b,c)$.
This is so as menu $(c-b,b,c)$ obtains at least as much revenue as menu $(a,b,c)$ for any valuations with $v_2\geq b$,
	and for any $v_2<b$, due to independence, the contribution to the expected gain of the event $v_1\geq c-b$ is at least as large as the contributed expected loss of the event  $v_1 \in [a, c-b)$.
\end{proof}

Hart and Reny~\cite[Corollary 9]{HR15-nonmon} have shown that submodular menus are revenue-monotone, i.e. when facing a higher (stochastically-dominating) valuation distribution, their revenue can never decrease.
The following corollary thus follows immediately from Theorem~\ref{thm:submodular}.

\begin{corollary}[Corollary \ref{cor:intro-rev-mon} restated]\label{cor:mon}
For every two distributions $\DD_1$ and $\DD_2$, and a single additive buyer with two items' valuations sampled from $\DD_1 \times \DD_2$, 
the revenue of any deterministic mechanism can be obtained by a  deterministic mechanism that is \emph{revenue-monotone}.
\end{corollary}

\subsection{\srev~vs \drev}
From Corollary \ref{cor:mon} we are able to determine exactly the maximum possible gap between the optimal deterministic auction
and the optimal revenue from selling the two items separately.  
\begin{theorem}[Theorem~\ref{thm:intro-1.278} restated]\label{thm:rev-srev}
Let $w\approx 1.278$ be the solution of the equation $(w-1)\cdot e^w = 1$.
For every two distributions $\DD_1$ and $\DD_2$ it holds 
that $\drev(\DD_1 \times \DD_2) \le w \cdot \srev(\DD_1 \times \DD_2)$.
That is, any revenue-optimal deterministic mechanism for selling two items with independent distributions obtains at most $w$-times more revenue than selling each item separately.
Furthermore, this is tight even for IID item distributions.
\end{theorem}
\ignore{
\section{Missing proofs from Section \ref{sec:main}}
\subsection{\srev~vs \drev}
\label{app:srev-drev}
From Corollary \ref{cor:mon} we are able to determine exactly the maximum possible gap between the optimal deterministic auction
and the optimal revenue from selling the two items separately.  
\begin{theorem}[Theorem~\ref{thm:intro-1.278} restated] 
	Let $w\approx 1.278$ be the solution of the equation $(w-1)\cdot e^w = 1$.
	For every two distributions $\DD_1$ and $\DD_2$ it holds that $\drev(\DD_1 \times \DD_2) \le w \cdot \srev(\DD_1 \times \DD_2)$.
	That is, any revenue-optimal deterministic mechanism for selling two items with independent distributions obtains at most $w$-times more revenue than selling each item separately.
	Furthermore, this is tight even for IID item distributions.
\end{theorem}
}

The tightness of our result follows from a result of \cite{HartN12}, that shows that for any ``equal revenue distribution'' $ER$, it holds that
$\drev(ER \times ER) = w \cdot \srev(ER \times ER)$. We next formally define equal revenue distributions - distributions such that any price that generates positive revenue, generates the same revenue.

\begin{definition}[Equal Revenue Distribution]
	For $r>0$, let the  {\em $r$-equal revenue distribution}, denoted by $\ERr{r}$, be the single-item valuation distribution whose cdf satisfies:
	$$ \Pr_{v \sim \ERr{r}}[v \geq p] = \min\left\{1, r/p\right\}.$$
\end{definition}

We next show that the revenue-optimal deterministic mechanism for selling two items whose valuations are drawn from independent (but possibly different) equal revenue distributions always sells both items together (i.e. $\drev = \brev$ for two independent equal revenue distributions).

\begin{lemma}\label{lem:eq-rev}
	For any $r_1,r_2>0$ it holds that 
	$$\drev (\ERr{r_1}\times \ERr{r_2})= \brev (\ERr{r_1}\times \ERr{r_2})$$	
\end{lemma}
\begin{proof}
	

	Fix a menu $m=(a,b,c)$ and assume WLOG that $a\leq b\leq c$. To prove the claim we show that any revenue that can be obtained by a deterministic menu, 
	can also be obtain by a menu in which the buyer never strictly prefers to buy a single item (over buying the pair or getting nothing). 
	A sufficient condition for such a menu is that $a=b=c$. 
	Consider any deterministic menu $m=(a,b,c)$ such that $a<c$, that is, it sells item $1$ for a cheaper price $a$ than the price $c$ charged for the bundle. 
	We will show that increasing the price of item $1$ to $c$ can only increase the revenue (the same arguments will clearly hold for the second item as well).
	That is, the menu $m'=(c,b,c)$ generates at least as much revenue as the menu $m=(a,b,c)$. 
	
	For $v_2\geq c-a$ the allocation and payments for both menus $m$ and $m'$ are the same, as it is never the case that the first item is bought alone, see Figure \ref{fig:DREV-BREV-ER} for an illustration. 
	We next argue about the case that $v_2<c-a$. 
	For any such case, 
	it holds that for any fixed $v_2$ the revenue with menu $m$ is $r_1$ (since item 2 is never sold and $v_1$ is distributed according to the equal revenue distribution $\ERr{r_1}$). 
	On the other hand, for $m'$, for any such $v_2< c-a$ the buyer either gets nothing or buys the pair, and the pair is sold in the event that $v_1+v_2\geq c$ which happen with probability at least as large as the probability that $v_1\geq c$. As selling item $1$ at price $c$ alone generates revenue $r_1$, the menu $m'$ which sells with at least as high probability, generates at least as high revenue. 
	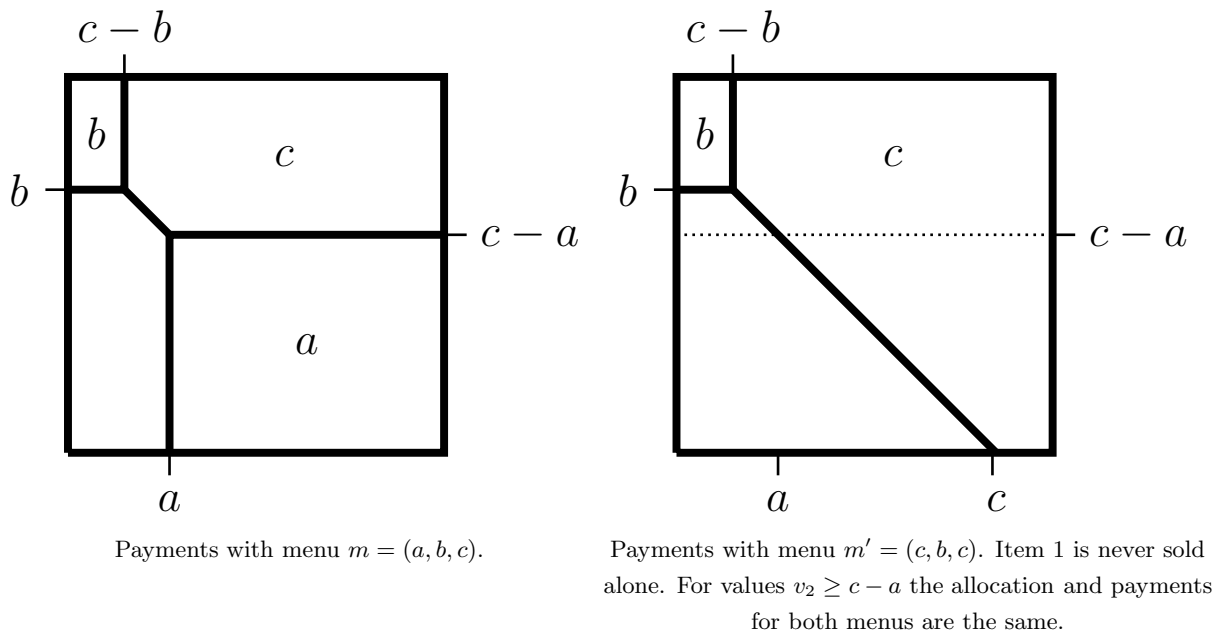
\begin{figure} \label{DREV-BREV-ER}
		\caption{DREV equals BREV for ER distributions}\label{fig:DREV-BREV-ER}
		\begin{minipage}[t]{0.49\columnwidth}
			\begin{center}
				\ifdefined\PAPER
\else

\documentclass[10pt]{article}
\usepackage{tikz, ifthen,etoolbox,color}
\usepackage{calc}
\usetikzlibrary{arrows.meta}
\newcommand{\ignore}[1]{}%

\begin{document}
\pagestyle{empty}

\fi

\ifdefined\SCALEA
\else
\def\SCALEA{0.07}
\fi
\begin{tikzpicture} [scale = \SCALEA]
\tikzstyle{every node}=[font=\LARGE]
\usetikzlibrary{calc}

\def\a{27} 
\def\b{70} 
\def\c{85} 

\def\width{3} 
\def\shift{1.5} 

\ifdefined\CHUPCHIK
\else
\def\CHUPCHIK{6}
\def\cwidth{1}
\fi

\draw  (0,0) -- (0,100) -- (100,100)-- (100,0) -- (0,0) [line width = \width];

\draw  (0,\b) -- (\c-\b,\b) -- (\c-\b,100)  [line width = \width];
\draw  (\a,0) -- (\a,\c-\a) -- (100,\c-\a)  [line width = \width];
\draw  (\a,\c-\a) -- (\c-\b,\b)  [line width = \width];

\draw  (-\CHUPCHIK,\b) -- (0, \b) [line width = \cwidth]; 
\draw  (\a, 0) -- (\a, -\CHUPCHIK) [line width = \cwidth]; 
\draw  (\c-\b,100+\CHUPCHIK) -- (\c-\b,100) [line width = \cwidth]; 
\draw  (100,\c-\a) -- (100+\CHUPCHIK,\c-\a)  [line width = \cwidth]; 

\draw (0.5*\a+50, 0.5*\c-0.5*\a) node {$a$};
\draw (0.5*\c-0.5*\b, 0.5*\b+50) node {$b$};
\draw (0.5*\c-0.5*\b+50, 0.5*\c-0.5*\a+50) node {$c$};

\draw (\a, -7-\CHUPCHIK) node {$a$};
\draw (-7-\CHUPCHIK, \b) node {$b$};
\draw (\c-\b, 107+\CHUPCHIK) node {$c-b$};
\draw (101+\CHUPCHIK,\c-\a) node [right] {$c-a$};

\end{tikzpicture}
\ifdefined\PAPER
\else
\end{document}
\fi 
				
				{\FIGURECAPTION Payments with menu $m=(a,b,c)$. }
			\end{center}
		\end{minipage}
		\begin{minipage}[t]{0.49\columnwidth}
			\begin{center}
				\ifdefined\PAPER
\else

\documentclass[10pt]{article}
\usepackage{tikz, ifthen,etoolbox,color}
\usepackage{calc}
\usetikzlibrary{arrows.meta}
\newcommand{\ignore}[1]{}%

\begin{document}
\pagestyle{empty}

\fi

\ifdefined\SCALEA
\else
\def\SCALEA{0.07}
\fi
\begin{tikzpicture} [scale = \SCALEA]
\tikzstyle{every node}=[font=\LARGE]
\usetikzlibrary{calc}

\def\a{27} 
\def\b{70} 
\def\c{85} 

\def\width{3} 
\def\shift{1.5} 

\ifdefined\CHUPCHIK
\else
\def\CHUPCHIK{6}
\def\cwidth{1}
\fi

\draw  (0,0) -- (0,100) -- (100,100)-- (100,0) -- (0,0) [line width = \width];

\draw  (0,\b) -- (\c-\b,\b) -- (\c-\b,100)  [line width = \width];
\draw  (0,\c-\a) -- (100,\c-\a)  [line width = \width/3, dotted];
\draw  (\c,0) -- (\c-\b,\b)  [line width = \width];

\draw  (-\CHUPCHIK,\b) -- (0, \b) [line width = \cwidth]; 
\draw  (\a, 0) -- (\a, -\CHUPCHIK) [line width = \cwidth]; 
\draw  (\c-\b,100+\CHUPCHIK) -- (\c-\b,100) [line width = \cwidth]; 
\draw  (100,\c-\a) -- (100+\CHUPCHIK,\c-\a)  [line width = \cwidth]; 
\draw  (\c  - 1, 0) -- (\c - 1, -\CHUPCHIK) [line width = \cwidth]; 

\draw (0.5*\c-0.5*\b, 0.5*\b+50) node {$b$};
\draw (0.5*\c-0.5*\b+50, 0.5*\c-0.5*\a+50) node {$c$};

\draw (\a, -7-\CHUPCHIK) node {$a$};
\draw (-7-\CHUPCHIK, \b) node {$b$};
\draw (\c-\b, 107+\CHUPCHIK) node {$c-b$};
\draw (101+\CHUPCHIK,\c-\a) node [right] {$c-a$};
\draw (\c, -7-\CHUPCHIK) node {$c$};

\end{tikzpicture}
\ifdefined\PAPER
\else
\end{document}
\fi
				
				{\FIGURECAPTION Payments with menu $m'=(c,b,c)$. Item $1$ is never sold alone. For values $v_2\geq c-a$ the allocation and payments for both menus are the same.}
			\end{center}
		\end{minipage}\hfill{}%
	\end{figure}
\end{proof}

We use the next Lemma from \cite{HartN12} in the proof of Theorem~\ref{thm:rev-srev} below.
\begin{lemma}[{\cite[Lemma 13]{HartN12}}] \label{lem:brev_srev}
	For every two distributions $\DD_1$ and $\DD_2$,
	$$\brev(\DD_1 \times \DD_2) \le w \cdot \srev(\DD_1 \times \DD_2)$$
	Furthermore, for some distributions this holds as equality (the result is tight).
\end{lemma}

We are now ready to prove Theorem~\ref{thm:rev-srev}.

\begin{proof}[Proof of Theorem~\ref{thm:rev-srev}]
	For $i = 1,2$ let $r_i = \rev(\DD_i)$ be the optimal revenue from selling item $i$ that is sampled from the distribution $\DD_i$.
	We note that any item value distribution $\FF$ for which the optimal revenue is $r$, is stochastically dominated by the distribution $\ERr{r}$.
	This implies, by Corollary \ref{cor:mon}, that $\drev(\DD_1 \times \DD_2)  \leq \drev(\ERr{r_1}\times \ERr{r_2})$. 
	We use this as the first step in our proof:
	\begin{align*}
	\drev(\DD_1 \times \DD_2) 
	& \leq \drev(\ERr{r_1}\times \ERr{r_2}) && \text{(\drev~is monotone by Cor.~\ref{cor:mon})} \\
	& \leq \brev(\ERr{r_1}\times \ERr{r_2}) && \text{(Lemma~\ref{lem:eq-rev})} \\
	& \leq w \cdot \srev(\ERr{r_1}\times \ERr{r_2}) && \text{(Lemma~\ref{lem:brev_srev})} \\
	& = w \cdot (r_1+r_2) && \text{(Def. of SRev, ER distributions)}\\
	& = w \cdot \srev(\DD_1 \times \DD_2)  && \text{(Def. of $r_1,r_2$)}.
	\end{align*}
\end{proof}

\section{Deterministic Pricing for Two IID Items}\label{sec:iid}
Assume that the two items are ex-ante symmetric, 
will a deterministic seller lose revenue by restricting her menu to be symmetric ($p_1=p_2$)?
For independent items, symmetric items means that they are sampled IID.
We show that the optimum revenue by deterministic mechanisms for selling two IID items can be achieved by \emph{symmetric} deterministic mechanisms.
We later show that the symmetry of the joint distribution of two IID items is \emph{not} sufficient to prove this result - for \emph{correlated} symmetric distribution the result does not hold. We demonstrate this by presenting an example in which an asymmetric deterministic mechanism  generates higher revenue than the best symmetric deterministic mechanism (Example~\ref{ex:correlated_symmetric}).    

We prove that for two IID items restricting to symmetric menus does not decrease the revenue.
The proof is similar in spirit to the proof of Theorem \ref{thm:submodular} but is much more involved and delicate%
\ifFULL
.
\else
~(see Appendix \ref{app:iid}).  
\fi
   
\begin{theorem}[Theorem~\ref{thm:intro-symmetric} restated]\label{thm:symmetric}
	For any value distribution $\FF$, and a single additive buyer with two items' valuations sampled from $\FF\times \FF$, 
	the revenue of any deterministic mechanism can be obtained by a deterministic mechanism that is \emph{symmetric}.
\end{theorem}	

\ifFULL

\ifFULL
\else
In this section we prove Theorem~\ref{thm:intro-symmetric}. We first restate the theorem. 
\begin{theorem}[Theorem~\ref{thm:intro-symmetric} restated]
	For any value distribution $\FF$, and a single additive buyer with two items' valuations sampled from $\FF\times \FF$, 
	the revenue of any deterministic mechanism can be obtained by a deterministic mechanism that is \emph{symmetric}.
\end{theorem}	
\fi

\begin{proof}	
	Given an asymmetric deterministic menu (that prices the two items differently), we construct a symmetric menu that obtains at least as much revenue. 
	
	Let $a,b,c$ denote the prices that the asymmetric menu charges for item $1$, item $2$, and the bundle, respectively. 
	Wlog, we can assume that $a < b \leq c$. Furthermore, by Theorem~\ref{thm:submodular}, we can assume wlog that $c \leq a+b$.
	It follows immediately that also $c \leq 2b$, but we do not know whether $c > 2a$. Our proof proceeds by analyzing separately the two possible cases.
	
	\subsubsection*{The case where $c \leq 2a$}
	
	The  case when $c \leq 2a$ is the easy one. 
	We show below that in this case, the symmetric menus $(a,a,c)$ and $(b,b,c)$ (i.e. charge both items $a$, respectively $b$) perform on average exactly as well as the original asymmetric menu $(a,b,c)$. Thus at least one of those symmetric menus has to yield at least as much revenue as the asymmetric menu $(a,b,c)$. 
	
	To prove that menus $(a,a,c)$ and $(b,b,c)$ yield as much revenue, on average, as the asymmetric menu, we partition  the valuation space across the line $v_1 = v_2$. 
 For $v_1 \geq v_2$, we argue that a buyer facing menus $(a,a,c)$ or $(a,b,c)$ never buys item $2$ by itself (without item $1$).
	For menu $(a,a,c)$, since $v_1\geq v_2$ and $c\leq 2a$, if $v_2\geq a$ then $v_1+v_2-c\geq v_1-a+v_2-a\geq v_2-a$ and thus whenever item $2$ gives non-negative utility, the pair is preferred to buying item $2$ by itself. 
	For menu $(a,b,c)$, $v_1\geq v_2$ and $a<b$ implies $v_1-a>v_2-b$, thus buying item $1$ by itself is preferred to buying item $2$ by itself. 
 Similarly, for $v_1 < v_2$, a buyer facing menus $(a,b,c)$ or $(b,b,c)$ never buys item $1$ by itself, so they yield the same revenue.
For menu $(a,b,c)$, since $v_1< v_2$ and $c\leq 2a$, if $v_1\geq a$ then $v_2>a\geq c-a$ and thus 
$v_1+v_2-c\geq v_1-a$, so whenever item $1$ gives non-negative utility, the pair is preferred to buying item $1$ by itself. 
For menu $(b,b,c)$, $v_1< v_2$ implies $v_1-b<v_2-b$, thus buying item $1$ by itself is preferred to buying item $2$ by itself.

	
	By symmetry, menus $(a,b,c)$ and $(b,a,c)$ yield the same revenue;
	$(b,a,c)$ is identical to $(a,a,c)$ on $v_1 < v_2$, and identical to $(b,b,c)$ on $v_1 \geq v_2$.
	Therefore,
	\begin{gather}
	\rev(a,a,c) + \rev(b,b,c) = \rev(a,b,c) + \rev(b,a,c) = 2\rev(a,b,c),
	\end{gather}
	where $\rev(x,y,z)$ denotes the expected revenue from menu $(x,y,z)$ for the given distribution.
	See also Figure~\ref{fig:c<2a}.
	Notice that this argument also works for correlated symmetric distributions.
	
	\begin{figure}
		\caption{Symmetric menus are optimal (Case $c \leq 2a$)}\label{fig:c<2a}
		\begin{center}
			\ifdefined\PAPER
\else

\documentclass[10pt]{article}
\usepackage{tikz, ifthen,etoolbox,color}
\usepackage{calc}
\usetikzlibrary{arrows.meta}
\newcommand{\ignore}[1]{}%

\begin{document}
\pagestyle{empty}

\fi

\begin{tikzpicture} [scale = 0.07]
\tikzstyle{every node}=[font=\LARGE]
\usetikzlibrary{calc}

\def\a{60} 
\def\b{75} 
\def\c{90} 

\def\width{3} 
\def\shift{1.5} 

\ifdefined\CHUPCHIK
\else
\def\CHUPCHIK{6}
\def\cwidth{1}
\fi

\draw  (0,0) -- (0,100) -- (100,100)-- (100,0) -- (0,0) [line width = \width];

\draw  (0,\b) -- (\c-\b,\b) -- (\c-\b,100)  [line width = \width];
\draw  (\a,0) -- (\a,\c-\a) -- (100,\c-\a)  [line width = \width];
\draw  (\a,\c-\a) -- (\c-\b,\b)  [line width = \width];

\draw  (-\CHUPCHIK,\b) -- (0, \b) [line width = \cwidth]; 
\draw  (\a, 0) -- (\a, -\CHUPCHIK) [line width = \cwidth]; 
\draw  (\c-\b,100+\CHUPCHIK) -- (\c-\b,100) [line width = \cwidth]; 
\draw  (100,\c-\a) -- (100+\CHUPCHIK,\c-\a)  [line width = \cwidth]; 

\draw  (0,\a+\shift) -- (\c-\a+\shift,\a+\shift) -- (\c-\a+\shift,100)  [line width = \width, color = brown];
\draw  (\a+\shift,0) -- (\a+\shift,\c-\a+\shift) -- (100,\c-\a+\shift)  [line width = \width, color = brown];
\draw  (\a+\shift,\c-\a+\shift) -- (\c-\a+\shift,\a+\shift)  [line width = \width, color = brown];

\draw  (0,\b-\shift) -- (\c-\b-\shift,\b-\shift) -- (\c-\b-\shift,100)  [line width = \width, color = green];
\draw  (\b-\shift,0) -- (\b-\shift,\c-\b-\shift) -- (100,\c-\b-\shift)  [line width = \width, color = green];
\draw  (\b-\shift,\c-\b-\shift) -- (\c-\b-\shift,\b-\shift)  [line width = \width, color = green];

\draw  (\b,0) -- (\b,\c-\b) -- (100, \c-\b)  [line width = \width, dotted];
\draw  (0,\a) -- (\c-\a,\a) -- (\c-\a, 100)  [line width = \width, dotted];
\draw  (\c-\a,\a) -- (\b,\c-\b,)  [line width = \width, dotted];


\draw (\a, -7-\CHUPCHIK) node {$a$};
\draw (-7-\CHUPCHIK, \b) node {$b$};
\draw (\c-\b, 107+\CHUPCHIK) node {$c-b$};
\draw (114+\CHUPCHIK, \c-\a) node {$c-a$};

\end{tikzpicture}

\ifdefined\PAPER
\else
\end{document}
\fi
			{\FIGURECAPTION Incentive constraints induced by menus $(a,b,c)$, $(b,a,c)$ (dotted), {\color{green}$(b,b,c)$}, and {\color{brown}$(a,a,c)$}.}
		\end{center}
	\end{figure}
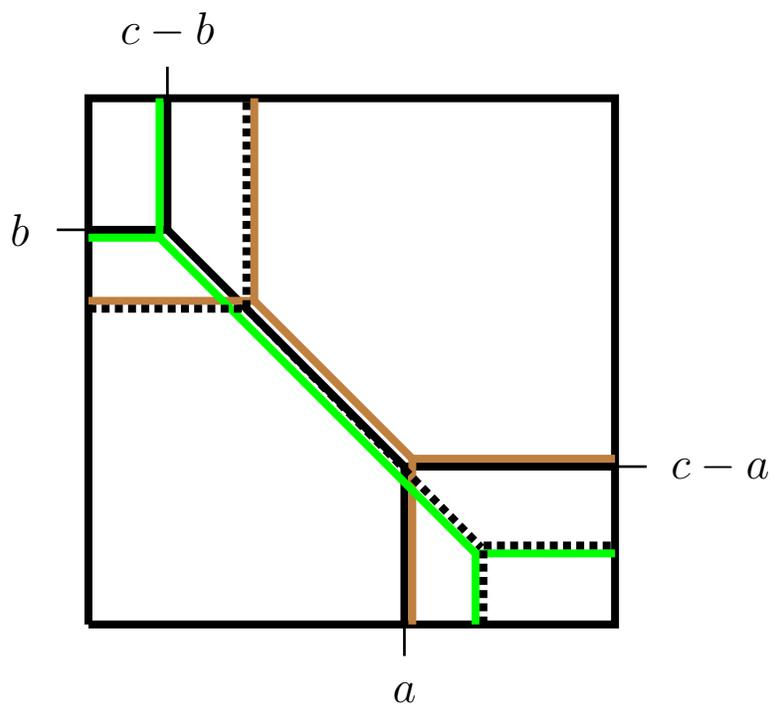
	
	\subsubsection*{The case where $c > 2a$}
	
	We consider three alternative symmetric menus: $(a,a,2a)$, $(b,b,c)$, and $(c-a, c-a, 2c-2a)$.%
	\footnote{Informally, we expect to have probability mass at $v_i \approx c-a$ (for $i = 1,2$) because this is where the buyer switches from item $1$ to the bundle in the asymmetric menu.}
%
	We show that at least one of them yields at least as much revenue as the asymmetric menu. 
	Specifically, we consider the following two cases:
	\begin{itemize}
		\item When 
		\begin{gather}\label{eq:green_red}
		2a \cdot \Pr \big[a \leq v_2 < c-a\big] \geq (c-2a) \cdot \Pr \big[v_2 \geq c-a\big],
		\end{gather} 
		we show that 
		\begin{gather*}
		\rev(b,b,c) + \rev(a,a,2a) \geq \rev(a,b,c) + \rev(b,a,c) = 2\rev(a,b,c).
		\end{gather*}
		thus one of the menus $(b,b,c)$ or $(a,a,2a)$ has revenue as least as high as the menu $(a,b,c)$.
		\item When 
		\begin{gather}\label{eq:green_blue}
		2a \cdot \Pr \big[a \leq v_2 < c-a\big] \leq (c-2a) \cdot \Pr \big[v_2 \geq c-a\big],
		\end{gather}
		we show that 
		\begin{gather*}
		\rev(b,b,c) + \rev(c-a, c-a, 2c-2a) \geq \rev(a,b,c) + \rev(b,a,c) = 2\rev(a,b,c).
		\end{gather*}
		thus one of the menus $(b,b,c)$ or $(c-a, c-a, 2c-2a)$ has revenue as least as high as the menu $(a,b,c)$.
	\end{itemize}

	Both proofs proceed as follows. First, we consider the partition of the valuation space induced by the incentive constraints of the respective four mechanisms (we consider a total of five menus, but we only need four for each case). Then, for each region of the space, we sum the payments to the seller in the symmetric menus, and subtract the payments in the asymmetric menus. 
	At this point, we still have some regions with negative and positive contributions.
	We now use the fact that the two items are symmetrically 
	distributed to observe that the revenue contribution of one region of valuations is exactly equal to the contribution of the symmetric region (when reflecting over the $v_1=v_2$ line), and we rearrange the revenue contributions without changing the total revenue using this fact. 
	Finally, we use the inequality on the single dimensional distributions (Inequality (\ref{eq:green_blue}) or (\ref{eq:green_red})) to show that the positive contributions outweigh the negative contributions.
	We note that this last argument is only true for the IID case and it fails for correlated distribution -- and indeed the claim is false for such distributions, see Example~\ref{ex:correlated_symmetric}.
	See Figures~\ref{fig:c>2a_I} and~\ref{fig:c>2a_I_green_blue} for more details.
	
	\begin{figure}
		\caption{Symmetric menus are optimal (Case $c > 2a$): the menus}\label{fig:c>2a_I}
		\vspace{1.5cm}
		\hspace*{\fill}
		\begin{minipage}[t]{0.49\columnwidth}
			\begin{center}
				\ifdefined\PAPER
\else

\documentclass[10pt]{article}
\usepackage{tikz, ifthen,etoolbox,color}
\usepackage{calc}
\usetikzlibrary{arrows.meta}
\newcommand{\ignore}[1]{}%

\begin{document}
\pagestyle{empty}
\fi
\ifdefined\SCALEA
\else
\def\SCALEA{0.07}
\fi
\begin{tikzpicture} [scale = \SCALEA]

\tikzstyle{every node}=[font=\LARGE]
\usetikzlibrary{calc}

\def\a{27} 
\def\b{70} 
\def\c{85} 

\def\width{3} 
\def\shift{1.5} 

\ifdefined\CHUPCHIK
\else
\def\CHUPCHIK{6}
\def\cwidth{1}
\fi

\draw  (0,0) -- (0,100) -- (100,100)-- (100,0) -- (0,0) [line width = \width];

\draw  (0,\b-\shift) -- (\c-\b-\shift,\b-\shift) -- (\c-\b-\shift,100)  [line width = \width, color = green];
\draw  (\b,0) -- (\b,\c-\b) -- (100,\c-\b)  [line width = \width, color = green];
\draw  (\b,\c-\b) -- (\c-\b-\shift,\b-\shift)  [line width = \width, color = green];

\draw (0.5*\b+50, 0.5*\c-0.5*\b) node [color = green] {$b$};
\draw (0.5*\c-0.5*\b, 0.5*\b+50) node [color = green] {$b$};
\draw (0.5*\c-0.5*\b+50, 0.5*\c-0.5*\b+50) node [color = green] {$c$};

\draw  (-\CHUPCHIK,\b) -- (0, \b) [line width = \cwidth]; 
\draw  (\c-\b,100+\CHUPCHIK) -- (\c-\b,100) [line width = \cwidth]; 

\draw (\a, -7-\CHUPCHIK) node [white] {$a$};
\draw (-7-\CHUPCHIK, \b) node {$b$};
\draw (\c-\b, 107+\CHUPCHIK) node {$c-b$};

\end{tikzpicture}
\ifdefined\PAPER
\else
\end{document}
\fi
				
				{\FIGURECAPTION Payments to seller with menu {\color{green}$(b,b,c)$}.
				Note that the menu $(b,b,c)$ is submodular as $2b\geq a+b\geq c$, so this plot as a submodular menu is correct (see Figure~\ref{two-types} for the difference in plots of submodular and supermodular menus).}
			\end{center}
		\end{minipage}\hspace*{\fill}
		\begin{minipage}[t]{0.49\columnwidth}
			\begin{center}
				\ifdefined\PAPER
\else

\documentclass[10pt]{article}
\usepackage{tikz, ifthen,etoolbox,color}
\usepackage{calc}
\usetikzlibrary{arrows.meta}
\newcommand{\ignore}[1]{}%

\begin{document}
\pagestyle{empty}
\fi
\ifdefined\SCALEA
\else
\def\SCALEA{0.07}
\fi
\begin{tikzpicture} [scale = \SCALEA]
\tikzstyle{every node}=[font=\LARGE]
\usetikzlibrary{calc}

\def\a{27} 
\def\b{70} 
\def\c{85} 

\def\width{3} 
\def\shift{1.5} 

\ifdefined\CHUPCHIK
\else
\def\CHUPCHIK{6}
\def\cwidth{1}
\fi

\draw  (0,0) -- (0,100) -- (100,100)-- (100,0) -- (0,0) [line width = \width];

\draw  (\c-\a,0) -- (\c-\a,100)  [line width = \width, color = blue];
\draw  (0,\c-\a+\shift) -- (100,\c-\a+\shift)  [line width = \width, color = blue];

\draw (0.5*\c-0.5*\a+50, 0.5*\c-0.5*\a) node [color = blue] {$c-a$};
\draw (0.5*\c-0.5*\a, 0.5*\c-0.5*\a+50) node [color = blue] {$c-a$};
\draw (0.5*\c-0.5*\a+50, 0.5*\c-0.5*\a+50) node [color = blue] {$2c-2a$};

\draw (\c-\a, -7-\CHUPCHIK) node {$c-a$};
\draw (-2-\CHUPCHIK, \c-\a) node [left] {$c-a$};

\draw  (0,\c-\a) -- (0-\CHUPCHIK,\c-\a)  [line width = \cwidth]; 
\draw  (\c-\a,0) -- (\c-\a,0-\CHUPCHIK)  [line width = \cwidth]; 

\end{tikzpicture}
\ifdefined\PAPER
\else
\end{document}
\fi
				
				{\FIGURECAPTION Payments to seller with menu {\color{blue}$(c-a,c-a,2c-2a)$}.}
			\end{center}
		\end{minipage}\hfill{}%
		
		\vspace{1.5cm}
		\hspace*{\fill}
		
		\begin{minipage}[t]{0.49\columnwidth}
			\begin{center}
				\ifdefined\PAPER
\else

\documentclass[10pt]{article}
\usepackage{tikz, ifthen,etoolbox,color}
\usepackage{calc}
\usetikzlibrary{arrows.meta}
\newcommand{\ignore}[1]{}%

\begin{document}
\pagestyle{empty}
\fi
\ifdefined\SCALEA
\else
\def\SCALEA{0.07}
\fi
\begin{tikzpicture} [scale = \SCALEA]
\tikzstyle{every node}=[font=\LARGE]
\usetikzlibrary{calc}

\def\a{27} 
\def\b{70} 
\def\c{85} 

\def\width{3} 
\def\shift{0} 

\ifdefined\CHUPCHIK
\else
\def\CHUPCHIK{6}
\def\cwidth{1}
\fi

\draw  (0,0) -- (0,100) -- (100,100)-- (100,0) -- (0,0) [line width = \width];


\draw  (\a+\shift,0) -- (\a+\shift,100)  [line width = \width, color = red];
\draw  (0,\a) -- (100,\a)  [line width = \width, color = red];

\draw (0.5*\a+50, 0.5*\a) node [color = red] {$a$};
\draw (0.5*\a, 0.5*\a+50) node [color = red] {$a$};
\draw (0.5*\a+50, 0.5*\a+50) node [color = red] {$2a$};

\draw (\a, -7-\CHUPCHIK) node {$a$};
\draw (-7-\CHUPCHIK, \a) node {$a$};

\draw  (0, \a) -- (-\CHUPCHIK, \a) [line width = \cwidth]; 
\draw  (\a, 0) -- (\a, -\CHUPCHIK) [line width = \cwidth]; 

\end{tikzpicture}
\ifdefined\PAPER
\else
\end{document}
\fi
				
				{\FIGURECAPTION Payments to seller with menu {\color{red}$(a,a,2a)$}.}
			\end{center}
		\end{minipage}\hspace*{\fill}
		\begin{minipage}[t]{0.49\columnwidth}
			\tikzstyle{every node}=[font=\scriptsize]
			\begin{center}
				\ifdefined\PAPER
\else

\documentclass[10pt]{article}
\usepackage{tikz, ifthen,etoolbox,color}
\usepackage{calc}
\usetikzlibrary{arrows.meta}
\newcommand{\ignore}[1]{}%

\begin{document}
\pagestyle{empty}
\fi
\ifdefined\SCALEB
\else
\def\SCALEB{0.07}
\fi
\begin{tikzpicture} [scale = \SCALEB]

\usetikzlibrary{calc}

\def\a{27} 
\def\b{70} 
\def\c{85} 

\def\width{3} 
\def\shift{1.5} 

\ifdefined\CHUPCHIK
\else
\def\CHUPCHIK{6}
\def\cwidth{1}
\fi

\draw  (0,0) -- (0,100) -- (100,100)-- (100,0) -- (0,0) [line width = \width];

\draw  (0,\b) -- (\c-\b,\b) -- (\c-\b,100)  [line width = \width];
\draw  (\a,0) -- (\a,\c-\a) -- (100,\c-\a)  [line width = \width];
\draw  (\a,\c-\a) -- (\c-\b,\b)  [line width = \width];

\draw  (-\CHUPCHIK,\b) -- (0, \b) [line width = \cwidth]; 
\draw  (\a, 0) -- (\a, -\CHUPCHIK) [line width = \cwidth]; 
\draw  (\c-\b,100+\CHUPCHIK) -- (\c-\b,100) [line width = \cwidth]; 
\draw  (100,\c-\a) -- (100+\CHUPCHIK,\c-\a)  [line width = \cwidth]; 

\draw  (\b,0) -- (\b,\c-\b) -- (100, \c-\b)  [line width = \width, dotted];
\draw  (0,\a) -- (\c-\a,\a) -- (\c-\a, 100)  [line width = \width, dotted];
\draw  (\c-\a,\a) -- (\b,\c-\b,)  [line width = \width, dotted];

\draw (0.5*\a, 0.5*\c) node {$a$};
\draw (0.5*\c-0.5*\b, 0.5*\b+50) node {$a+b$};
\draw (0.5*\c, 0.5*\c) node {$2a$};
\draw (\c-0.5*\a-0.5*\b, 0.5*\c-0.5*\a + 50) node {$a+c$};
\draw (0.5*\c-0.5*\a+50, 0.5*\c-0.5*\a+50) node {$2c$};

\draw (0.5*\c, 0.5*\a) node {$a$};
\draw (0.5*\b+50, 0.5*\c-0.5*\b) node {$a+b$};
\draw (0.5*\c-0.5*\a + 50, \c-0.5*\a-0.5*\b) node {$a+c$};

\draw (\a, -7-\CHUPCHIK) node {$a$};
\draw (-2-\CHUPCHIK, \b) node [left] {$b$};
\draw (\c-\b, 107+\CHUPCHIK) node {$c-b$};
\draw (103+\CHUPCHIK,\c-\a) node [right] {$c-a$};

\end{tikzpicture}
\ifdefined\PAPER
\else
\end{document}
\fi
				
				{\FIGURECAPTION Sum of payments from menus $(a,b,c)$ and $(b,a,c)$ (dotted).%
				\footnote{See also the right plot in Figure~\ref{two-types}.}}
			\end{center}
		\end{minipage}\hfill{}%
		
	\end{figure}
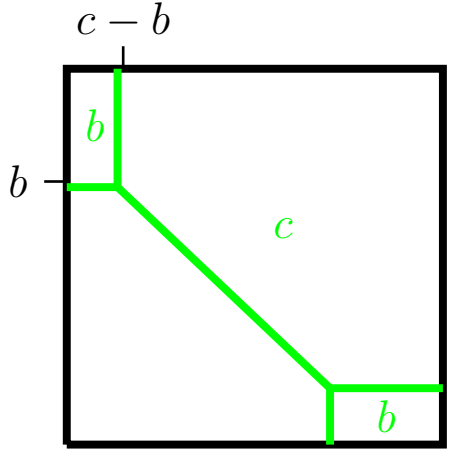

	\begin{figure}
		\caption{Symmetric menus are optimal (Case $c > 2a$): averaging arguments}\label{fig:c>2a_I_green_blue}
		\tikzstyle{every node}=[font=\scriptsize]
		\hspace*{\fill}

		\hspace*{\fill}
		\begin{minipage}[t]{\MINIPAGEWIDTH\columnwidth}
			\ifdefined\PAPER
\else

\documentclass[10pt]{article}
\usepackage{tikz, ifthen,etoolbox,color}
\usepackage{calc}
\usetikzlibrary{arrows.meta}
\newcommand{\ignore}[1]{}%

\begin{document}
\pagestyle{empty}
\fi
\ifdefined\SCALEB
\else
\def\SCALEB{0.07}
\fi
\begin{tikzpicture} [scale = \SCALEB]
\usetikzlibrary{calc}

\def\a{27} 
\def\b{70} 
\def\c{85} 

\def\width{3} 
\def\shift{1.5} 

\ifdefined\CHUPCHIK
\else
\def\CHUPCHIK{6}
\def\cwidth{1}
\fi

\draw  (0,0) -- (0,100) -- (100,100)-- (100,0) -- (0,0) [line width = \width];

\draw  (\a+\shift,0) -- (\a+\shift,100)  [line width = \width, color = red];
\draw  (0,\a) -- (100,\a)  [line width = \width, color = red];

\draw  (0,\b-\shift) -- (\c-\b-\shift,\b-\shift) -- (\c-\b-\shift,100)  [line width = \width, color = green];
\draw  (\b,0) -- (\b,\c-\b) -- (100,\c-\b)  [line width = \width, color = green];
\draw  (\b,\c-\b) -- (\c-\b-\shift,\b-\shift)  [line width = \width, color = green];

\draw  (0,\b) -- (\c-\b,\b) -- (\c-\b,100)  [line width = \width];
\draw  (\a,0) -- (\a,\c-\a) -- (100,\c-\a)  [line width = \width];
\draw  (\a,\c-\a) -- (\c-\b,\b)  [line width = \width];

\draw  (\b,0) -- (\b,\c-\b) -- (100, \c-\b)  [line width = \width, dotted];
\draw  (0,\a) -- (\c-\a,\a) -- (\c-\a, 100)  [line width = \width, dotted];
\draw  (\c-\a,\a) -- (\b,\c-\b,)  [line width = \width, dotted];

\draw  (-\CHUPCHIK,\b) -- (0, \b) [line width = \cwidth]; 
\draw  (\a, 0) -- (\a, -\CHUPCHIK) [line width = \cwidth]; 
\draw  (\c-\b,100+\CHUPCHIK) -- (\c-\b,100) [line width = \cwidth]; 
\draw  (100,\c-\a) -- (100+\CHUPCHIK,\c-\a)  [line width = \cwidth]; 

\draw (0.75*\c - 0.5*\a, 0.75*\c - 0.5*\a) node {$c$};
\draw (0.5*\c, 0.5*\c-0.5*\a + 50) node {$a$};
\draw (0.5*\c-0.5*\a+50, 0.5*\c-0.5*\a+50) node {$2a-c$};

\draw (0.5*\c-0.5*\a + 50, 0.5*\c) node {$a$};

\draw (\a, -7-\CHUPCHIK) node {$a$};
\draw (-2-\CHUPCHIK, \b) node [left] {$b$};
\draw (\c-\b, 107+\CHUPCHIK) node {$c-b$};
\draw (103+\CHUPCHIK,\c-\a) node [right] {$c-a$};

\end{tikzpicture}
\ifdefined\PAPER
\else
\end{document}
\fi
		\end{minipage}\hfill{}%
		\begin{minipage}[t]{\MINIPAGEWIDTH\columnwidth}
			\ifdefined\PAPER
\else

\documentclass[10pt]{article}
\usepackage{tikz, ifthen,etoolbox,color}
\usepackage{calc}
\usetikzlibrary{arrows.meta}
\newcommand{\ignore}[1]{}%

\begin{document}
\pagestyle{empty}
\fi
\ifdefined\SCALEB
\else
\def\SCALEB{0.07}
\fi
\begin{tikzpicture} [scale = \SCALEB]
\usetikzlibrary{calc}

\def\a{27} 
\def\b{70} 
\def\c{85} 

\def\width{3} 
\def\shift{1.5} 

\ifdefined\CHUPCHIK
\else
\def\CHUPCHIK{6}
\def\cwidth{1}
\fi

\draw  (0,0) -- (0,100) -- (100,100)-- (100,0) -- (0,0) [line width = \width];

\draw  (\a+\shift,0) -- (\a+\shift,100)  [line width = \width, color = red];
\draw  (0,\a) -- (100,\a)  [line width = \width, color = red];

\draw  (0,\b-\shift) -- (\c-\b-\shift,\b-\shift) -- (\c-\b-\shift,100)  [line width = \width, color = green];
\draw  (\b,0) -- (\b,\c-\b) -- (100,\c-\b)  [line width = \width, color = green];
\draw  (\b,\c-\b) -- (\c-\b-\shift,\b-\shift)  [line width = \width, color = green];

\draw  (0,\b) -- (\c-\b,\b) -- (\c-\b,100)  [line width = \width];
\draw  (\a,0) -- (\a,\c-\a) -- (100,\c-\a)  [line width = \width];
\draw  (\a,\c-\a) -- (\c-\b,\b)  [line width = \width];

\draw  (\b,0) -- (\b,\c-\b) -- (100, \c-\b)  [line width = \width, dotted];
\draw  (0,\a) -- (\c-\a,\a) -- (\c-\a, 100)  [line width = \width, dotted];
\draw  (\c-\a,\a) -- (\b,\c-\b,)  [line width = \width, dotted];

\draw  (-\CHUPCHIK,\b) -- (0, \b) [line width = \cwidth]; 
\draw  (\a, 0) -- (\a, -\CHUPCHIK) [line width = \cwidth]; 
\draw  (\c-\b,100+\CHUPCHIK) -- (\c-\b,100) [line width = \cwidth]; 
\draw  (100,\c-\a) -- (100+\CHUPCHIK,\c-\a)  [line width = \cwidth]; 

\draw (0.75*\c - 0.5*\a, 0.75*\c - 0.5*\a) node {$c$};
\draw (0.5*\c, 0.5*\c-0.5*\a + 50) node {$2a$};
\draw (0.5*\c-0.5*\a+50, 0.5*\c-0.5*\a+50) node {$2a-c$};


\draw (\a, -7-\CHUPCHIK) node {$a$};
\draw (-2-\CHUPCHIK, \b) node [left] {$b$};
\draw (\c-\b, 107+\CHUPCHIK) node {$c-b$};
\draw (103+\CHUPCHIK,\c-\a) node [right] {$c-a$};

\end{tikzpicture}
\ifdefined\PAPER
\else
\end{document}
\fi
		\end{minipage}\hspace*{\fill}
		
		{\FIGURECAPTION Menus $(a,b,c)$, $(b,a,c)$ (dotted), {\color{green}$(b,b,c)$}, and {\color{red}$(a,a,a)$}.
		
		On the left: the number inside each region represents the sum of contributions from {\color{green}$(b,b,c)$} and {\color{red}$(a,a,2a)$}, minus the contributions from $(a,b,c)$ and $(b,a,c)$.
		
		On the right: 
		we rearranged the payments, moving $a$ from the right-center region to the top-center region.
		By symmetry, the probability mass on those two regions is the same, so the total revenue does not change. Whenever \eqref{eq:green_red} holds, we have that the total contribution in each horizontal
		line (fixed $v_2$) is nonnegative.}
		
		\vspace{0.3cm}
		\begin{minipage}[t]{\MINIPAGEWIDTH\columnwidth}
			\begin{center}
				\ifdefined\PAPER
\else

\documentclass[10pt]{article}
\usepackage{tikz, ifthen,etoolbox,color}
\usepackage{calc}
\usetikzlibrary{arrows.meta}
\newcommand{\ignore}[1]{}%

\begin{document}
\pagestyle{empty}
\fi
\ifdefined\SCALEB
\else
\def\SCALEB{0.07}
\fi
\begin{tikzpicture} [scale = \SCALEB]
\usetikzlibrary{calc}
\usetikzlibrary{shapes.multipart}

\def\a{27} 
\def\b{70} 
\def\c{85} 

\def\width{3} 
\def\shift{1.5} 

\ifdefined\CHUPCHIK
\else
\def\CHUPCHIK{6}
\def\cwidth{1}
\fi

\draw  (0,0) -- (0,100) -- (100,100)-- (100,0) -- (0,0) [line width = \width];

\draw  (\c-\a,0) -- (\c-\a,100)  [line width = \width, color = blue];
\draw  (0,\c-\a+\shift) -- (100,\c-\a+\shift)  [line width = \width, color = blue];

\draw  (0,\b-\shift) -- (\c-\b-\shift,\b-\shift) -- (\c-\b-\shift,100)  [line width = \width, color = green];
\draw  (\b,0) -- (\b,\c-\b) -- (100,\c-\b)  [line width = \width, color = green];
\draw  (\b,\c-\b) -- (\c-\b-\shift,\b-\shift)  [line width = \width, color = green];

\draw  (0,\b) -- (\c-\b,\b) -- (\c-\b,100)  [line width = \width];
\draw  (\a,0) -- (\a,\c-\a) -- (100,\c-\a)  [line width = \width];
\draw  (\a,\c-\a) -- (\c-\b,\b)  [line width = \width];

\draw  (\b,0) -- (\b,\c-\b) -- (100, \c-\b)  [line width = \width, dotted];
\draw  (0,\a) -- (\c-\a,\a) -- (\c-\a, 100)  [line width = \width, dotted];
\draw  (\c-\a,\a) -- (\b,\c-\b,)  [line width = \width, dotted];

\draw  (-\CHUPCHIK,\b) -- (0, \b) [line width = \cwidth]; 
\draw  (\a, 0) -- (\a, -\CHUPCHIK) [line width = \cwidth]; 
\draw  (\c-\b,100+\CHUPCHIK) -- (\c-\b,100) [line width = \cwidth]; 
\draw  (0,\c-\a) -- (0-\CHUPCHIK,\c-\a)  [line width = \cwidth]; 

\draw (0.5*\a, 0.5*\c) node {$-a$};
\draw (0.5*\c-0.5*\b, 0.5*\b+50) node [rotate = 70] {$c$$-$$2a$};\draw (0.5*\c-0.5*\b, 0.5*\b+0.5*\c-0.5*\a) node {$c$$-$$2a$};
\draw (0.25*\c + 0.5*\a, 0.25*\c + 0.5*\a) node {$-2a$};
\draw (0.75*\c - 0.5*\a, 0.75*\c - 0.5*\a) node {$c$$-$$2a$};
\draw (0.5*\c, 0.5*\c-0.5*\a + 50) node {$c-2a$};
\draw (0.5*\c-0.5*\a+50, 0.5*\c-0.5*\a+50) node {$c-2a$};

\draw (0.5*\c, 0.5*\a) node {$-a$};
\draw (0.5*\b+50, 0.5*\c-0.5*\b) node {$c-2a$};
\draw (0.5*\b+0.5*\c-0.5*\a, 0.5*\c-0.5*\b) node [rotate = 70]{$c$$-$$2a$};\draw (0.5*\c-0.5*\a + 50, 0.5*\c) node {$c-2a$};

\draw (\a, -7-\CHUPCHIK) node {$a$};
\draw (-2-\CHUPCHIK, \b) node [left] {$b$};
\draw (\c-\b, 107+\CHUPCHIK) node {$c-b$};
\draw (-2-\CHUPCHIK,\c-\a) node [left] {$c-a$};

\end{tikzpicture}
\ifdefined\PAPER
\else
\end{document}
\fi
			\end{center}
		\end{minipage}\hspace*{\fill}
		\begin{minipage}[t]{\MINIPAGEWIDTH\columnwidth}
			\begin{center}
				\ifdefined\PAPER
\else

\documentclass[10pt]{article}
\usepackage{tikz, ifthen,etoolbox,color}
\usepackage{calc}
\usetikzlibrary{arrows.meta}
\newcommand{\ignore}[1]{}%

\begin{document}
\pagestyle{empty}
\fi
\ifdefined\SCALEB
\else
\def\SCALEB{0.07}
\fi
\begin{tikzpicture} [scale = \SCALEB]
\usetikzlibrary{calc}
\usetikzlibrary{shapes.multipart}

\def\a{27} 
\def\b{70} 
\def\c{85} 

\def\width{3} 
\def\shift{1.5} 

\ifdefined\CHUPCHIK
\else
\def\CHUPCHIK{6}
\def\cwidth{1}
\fi

\draw  (0,0) -- (0,100) -- (100,100)-- (100,0) -- (0,0) [line width = \width];

\draw  (\c-\a,0) -- (\c-\a,100)  [line width = \width, color = blue];
\draw  (0,\c-\a+\shift) -- (100,\c-\a+\shift)  [line width = \width, color = blue];

\draw  (0,\b-\shift) -- (\c-\b-\shift,\b-\shift) -- (\c-\b-\shift,100)  [line width = \width, color = green];
\draw  (\b,0) -- (\b,\c-\b) -- (100,\c-\b)  [line width = \width, color = green];
\draw  (\b,\c-\b) -- (\c-\b-\shift,\b-\shift)  [line width = \width, color = green];

\draw  (0,\b) -- (\c-\b,\b) -- (\c-\b,100)  [line width = \width];
\draw  (\a,0) -- (\a,\c-\a) -- (100,\c-\a)  [line width = \width];
\draw  (\a,\c-\a) -- (\c-\b,\b)  [line width = \width];

\draw  (\b,0) -- (\b,\c-\b) -- (100, \c-\b)  [line width = \width, dotted];
\draw  (0,\a) -- (\c-\a,\a) -- (\c-\a, 100)  [line width = \width, dotted];
\draw  (\c-\a,\a) -- (\b,\c-\b,)  [line width = \width, dotted];

\draw  (-\CHUPCHIK,\b) -- (0, \b) [line width = \cwidth]; 
\draw  (\a, 0) -- (\a, -\CHUPCHIK) [line width = \cwidth]; 
\draw  (\c-\b,100+\CHUPCHIK) -- (\c-\b,100) [line width = \cwidth]; 
\draw  (0,\c-\a) -- (0-\CHUPCHIK,\c-\a)  [line width = \cwidth]; 

\draw (0.5*\c-0.5*\b, 0.5*\b+50) node [rotate = 70] {$c$$-$$2a$};\draw (0.5*\c-0.5*\b, 0.5*\b+0.5*\c-0.5*\a) node {$c$$-$$2a$};
\draw (0.25*\c + 0.5*\a, 0.25*\c + 0.5*\a) node {$-2a$};
\draw (0.75*\c - 0.5*\a, 0.75*\c - 0.5*\a) node {$c$$-$$2a$};
\draw (0.5*\c, 0.5*\c-0.5*\a + 50) node {$c-2a$};
\draw (0.5*\c-0.5*\a+50, 0.5*\c-0.5*\a+50) node {$c-2a$};

\draw (0.5*\c, 0.5*\a) node {$-2a$};
\draw (0.5*\b+50, 0.5*\c-0.5*\b) node {$c-2a$};
\draw (0.5*\b+0.5*\c-0.5*\a, 0.5*\c-0.5*\b) node [rotate = 70]{$c$$-$$2a$};\draw (0.5*\c-0.5*\a + 50, 0.5*\c) node {$c-2a$};

\draw (\a, -7-\CHUPCHIK) node {$a$};
\draw (-2-\CHUPCHIK, \b) node [left] {$b$};
\draw (\c-\b, 107+\CHUPCHIK) node {$c-b$};
\draw (-2-\CHUPCHIK,\c-\a) node [left] {$c-a$};

\end{tikzpicture}
\ifdefined\PAPER
\else
\end{document}
\fi
			\end{center}
		\end{minipage}\hspace*{\fill}
		
		{\FIGURECAPTION  Menus $(a,b,c)$, $(b,a,c)$ (dotted), {\color{green}$(b,b,c)$}, and {\color{blue}$(c-a,c-a,2c-2a)$}. 
		
		On the left: 
		The number inside each region represents the sum of contributions from {\color{green}$(b,b,c)$} and {\color{blue}$(c-a,c-a,2c-2a)$}, minus the contributions from $(a,b,c)$ and $(b,a,c)$.
		
		On the right: We rearranged the payments, moving $a$ from the left-center region to the center-bottom region. 
		By symmetry, the probability mass on those two regions is the same, so the total revenue does not change.
		Notice that whenever \eqref{eq:green_blue} holds, we have that the total contribution in each  horizontal 
		line (fixed $v_2$) is nonnegative. }
		
	\end{figure}
\end{proof}





\fi

\section{More than Two Items}      
In this section we show that our results for two items 
do not extend even to three items, even when the values of the three items are sampled IID from a distribution with a finite support.
While for two independent items the revenue of a deterministic menu can be obtained with a submodular menu, for three IID items this is not true. 
Additionally, for three items symmetric menus are losing revenue compared to arbitrary deterministic menus. 

\begin{theorem}
	There exist a distribution $\FF$ with finite support such that for three items sampled IID from $\FF$ ($\DD= \FF\times \FF\times \FF$), the following holds\footnote{Note that since the support is finite, the optimal mechanism in each case is indeed obtained.}:
	The expected revenue of the optimal deterministic mechanism is strictly larger than both the expected revenue of the optimal \emph{symmetric} deterministic mechanism and of the optimal \emph{submodular} deterministic mechanism.	 
\end{theorem}

The theorem directly follows from the next example which further shows that imposing both symmetry and submodularity decreases the revenue even more than imposing only one of them.

\begin{example}[$n=3$] \label{ex:n=3}
	Let $\FF$ be following discrete distribution that samples values uniformly from the multi-set $[0, 1, 2, 2, 2, 2, 5, 6, 6, 6]$, that is: $Pr[v=0]=Pr[v=1]=Pr[v=5]=0.1$, $Pr[v=2]=0.4$ and $Pr[v=6]=0.3$.  
	It holds that:
	\begin{itemize}
		\item The expected revenue of the optimal deterministic mechanism is $6.293$.
		\item The expected revenue of the optimal \emph{symmetric} deterministic mechanism is $6.291$.
		\item The expected revenue of the optimal \emph{submodular} deterministic mechanism is $6.292$.
		\item The expected revenue of the optimal \emph{symmetric and submodular} deterministic mechanism is $6.288$.
	\end{itemize}
	First, observe that it is enough to only consider integer prices, bounded above by appropriate prices (e.g., not more than 6 for a single item, not more than 12 for 2 items, etc.). Using this fact, by exhaustive search we found the optimal mechanisms in each class. These optimal mechanisms, with prices $(p_1,p_2,p_3,p_{1,2},p_{1,3},p_{2,3},p_{1,2,3})$, are as follows: 
	\begin{itemize}
		\item $(6, 6, 6, 7, 7, 8, 9)$ is an optimal deterministic mechanism;
		\item $(6, 6, 6, 7, 7, 7, 9)$ is an optimal symmetric deterministic mechanism;
		\item $(5, 6, 6, 7, 7, 8, 9)$ is an optimal submodular deterministic mechanism; and
		\item $(5, 5, 5, 7, 7, 7, 9)$ is an optimal symmetric and submodular deterministic mechanism. 
			\qed
	\end{itemize}
\end{example}    

This example does not rule out, however, that {\em subadditive} deterministic mechanisms can get as much revenue as any other deterministic mechanism.
Resolving whether this is indeed the case is the main open problem we propose in this work:
\begin{oq}
	For $n > 2$ items with independent valuations, is it true that the revenue of any deterministic mechanism can be obtained by a deterministic mechanism that is {\em subadditive}?
\end{oq}

	Additionally, we have shown that for two items, revenue-monotone deterministic mechanisms can get as much revenue as any other mechanisms. We leave open the question if this is true for more than two items.
	\begin{oq}
		For $n>2$ items with independent valuations, 
		is it true that the revenue of any deterministic mechanism can be obtained by a deterministic mechanism that is {\em revenue monotone}?
	\end{oq}




\section{Correlated Valuations}
\label{sec:cor}
We next aim to understand to what extent the results for deterministic auction for selling two items with values sampled independently, extend to two item with values sampled from a \emph{correlated joint distribution}. We have seen that for two independent items restricting to submodular menu does not result in any revenue loss for the seller. We next show that for correlated items, the situation is very different - moving from submodular menu to an unrestricted one can increase the revenue of the seller, and that increase can be as large as almost 50\%. We also show that this is tight and the gain is never larger than 50\%.

\subsection{Price of Submodularity}
The following example shows that for two items with values sampled from a correlated distribution, the optimal deterministic auction can gain almost 50\% more revenue than the optimal deterministic submodular auction --- and this is true even for a symmetric correlated distribution.
\begin{example}[$\drev \geq (3/2-\varepsilon) \smdrev$] \label{ex:correlated-supermod}
	
	Fix a small $\varepsilon>0$. Consider two items with valuations drawn from the following distribution:
	$$ (v_1, v_2) = \begin{cases}
	(4,0) & \text{w.p. $1/2 - \varepsilon$}\\
	(0,4) & \text{w.p. $1/2 - \varepsilon$}\\
	(1/\varepsilon,1/\varepsilon) & \text{w.p. $2\varepsilon$}
	\end{cases}.$$
	
	There is a high-revenue deterministic auction that charges $4$ for each item or $1/\varepsilon$ for the grand bundle. Its revenue is $\drev \geq 4(1-2\varepsilon) + (1/\varepsilon)\cdot2\varepsilon = 6-O(\varepsilon)$.
	In contrast we show below that any 
	submodular auction has revenue at most $4+O(\varepsilon)$. 
	
	We consider three different cases:
	\begin{itemize}
		\item If the auction prices each of the two items at a price larger than $4$, the buyer does not buy anything for any valuation in her support except $(1/\varepsilon, 1/\varepsilon)$. The seller's revenue is at most the welfare from this event, $(1/\varepsilon+ 1/\varepsilon)\cdot 2\varepsilon = 4$.
		\item Suppose the auction charges at most $4$ for one of the items, and more than $4$ for the other item. Then the grand bundle never sells for price more than $1/\varepsilon + 4$ --- otherwise even with valuration $(1/\varepsilon, 1/\varepsilon)$ the buyer prefers buying the cheaper item over the grand bundle. Furthermore, the buyer never buys the more expensive item. Therefore the seller's revenue is bounded from above by $2$ from selling the cheaper item and  $2+O(\varepsilon)$ from selling the grand bundle.
		\item Finally, any submodular auction that charges at most $4$ for both items (and hence, by submodularity, at most $8$ for the grand bundle) can make at most $O(\varepsilon)$ revenue from the high valuation. 
	\end{itemize}
	\qed
\end{example}

The following lemma shows that the previous example is tight.


\begin{lemma}
	For any joint value distribution $\DD$ over two items it holds that 
	$$ \drev(\DD) \leq   (3/2) \cdot \smdrev(\DD).$$
\end{lemma}
\begin{proof}
	Throughout the proof we use $\rev(x,y,z)$ to denote the expected revenue from menu $(x,y,z)$ over the distribution $\DD$.
	Fix any deterministic menu $(p_1, p_2, p_{1,2})$. 
	Assume that the menu  $m=(p_1, p_2, p_{1,2})$ is not submodular, that is, $p_{1,2}>p_1+p_2$. 
	Let $A_{\emptyset}, A_1, A_2, A_{1,2}$ denote the events that, given prices $p_1$, $p_2$, and $p_{1,2}$, the buyer buys the empty set, only item $1$, etc. 
	Now, the revenue from the menu  $(p_1, p_2, p_{1,2})$ is given by:
	$$ \rev(p_1, p_2, p_{1,2}) = \Pr[A_1]p_1 + \Pr[A_2]p_2 + \Pr[A_{1,2}]p_{1,2}.$$
	
	Let $p=2p_{1,2}-p_1-p_2$. To complete the proof we will show that we can bound  $\rev(m)= \rev(p_1, p_2, p_{1,2})$ by half the revenue obtained by selling only the grand bundle at price $p$ (with the menu $(p, p, p)$) plus the revenue obtained by selling the two items separately at prices $p_1$ and $p_2$ (with the additive menu $(p_1,p_2, p_1+p_2)$). 
	This implies that $3/2$ times the better revenue of the two menus, is at least the revenue of the menu $m$.
	I.e. we show that 
	$$ \rev(m)
	\leq  \rev(p_1, p_2, p_1+p_2) + (1/2) \rev(p, p, p) \leq 
	(3/2) \cdot \max\{\rev(p_1, p_2, p_1+p_2), \rev(p, p, p)\},
	$$ 
	which is clearly sufficient to complete the proof as the menus $(p_1, p_2, p_1+p_2)$ and $(p, p, p)$ are both submodular.
	
	For any valuation $(v_1, v_2)$ for which the buyer buys the grand bundle (i.e. event $A_{1,2}$ occurs) with menu $(p_1,p_2, p_{1,2})$, it holds that he prefers the bundle over any item $i\in \{1,2\}$, thus $v_1 +v_2 - p_{1,2}\geq v_i-p_i$. So by summing these two up, in any such case it holds that  $v_1 +v_2 \geq 2 p_{1,2}-p_1-p_2 =p$. Thus with menu $(p, p, p)$, for any such valuation the buyer prefers buying the grand bundle over the empty set. Therefore,  
	$$\rev(p, p, p) \geq \Pr[A_{1,2}] (2p_{1,2}-p_1-p_2).$$
	
	Additionally, for the mechanism that offers each item $i$ separately for price $p_i$ (and the bundle for $p_1 + p_2$) it holds that for buyer's valuations in event $A_{1,2}$, the buyer buys both items since for any items $i\neq j\in \{1,2\}$ as  
	$v_1 +v_2 - p_{1,2}\geq v_i-p_i$ it holds that $v_j \geq p_{1,2} - p_i> p_j$ (by supermodularity of the menu $(p_1, p_2, p_{1,2})$), 
	thus 
	$$ \rev(p_1, p_2, p_1+p_2) \geq (\Pr[A_1] + \Pr[A_{1,2}])p_1 + (\Pr[A_2] + \Pr[A_{1,2}])p_2.$$
	
	Combining the above two inequalities: 
	
	\begin{align*}
	\rev(p_1, p_2, p_1+p_2) & + (1/2) \rev(p, p, p)  \\
	& \geq (\Pr[A_1] + \Pr[A_{1,2}])p_1 + (\Pr[A_2] + \Pr[A_{1,2}])p_2 + \Pr[A_{1,2}] (p_{1,2}-p_1 /2 -p_2 /2)\\
	& = (\Pr[A_1] + \Pr[A_{1,2}]/2)p_1 + (\Pr[A_2] + \Pr[A_{1,2}]/2)p_2 + \Pr[A_{1,2}]p_{1,2} \\
	& \geq \Pr[A_1]p_1 + \Pr[A_2]p_2 + \Pr[A_{1,2}]p_{1,2}\\
	& =  \rev(p_1, p_2, p_{1,2}).
	\end{align*}
	
\end{proof}

\subsection{Symmetric vs Asymmetric Menus}
    Consider a joint distribution that is symmetric. 
    Will a deterministic seller lose revenue by restricting her menu to be symmetric when items are correlated?
   %
%
	We next show that for correlated distributions over two items, the fact that the joint distribution is symmetric is {\em not} sufficient to ensure that there is an optimal deterministic menu that is symmetric. Moreover, there is a constant factor gap between the achievable revenues.

\ignore{
\subsection{Symmetric vs Asymmetric Menus}
}
\begin{example}[Asymmetric menus for symmetric distributions] \label{ex:correlated_symmetric}
	
	Fix a small $\varepsilon>0$ and consider the following	correlated distribution over two-item valuations:
	$$
	(v_1, v_2) \sim 
	\begin{cases}
	(1/\varepsilon^2, 1/\varepsilon^2) & \text{w.p. $\varepsilon^2$}\\
	(1/\varepsilon, 0) & \text{w.p. $\varepsilon/2$}\\
	(0, 1/\varepsilon) & \text{w.p. $\varepsilon/2$}\\
	(1,1) & \text{otherwise}
	\end{cases}
	$$
	There is a deterministic asymmetric menu with prices $(1, 1/\varepsilon, 1/\varepsilon^2)$ and it obtains revenue of $5/2 - O(\varepsilon)$.

	In contrast, we claim that any symmetric menu $(p, p, q)$ yields revenue at most $2+\varepsilon$.
	We consider three different cases, depending on the price $p$ of any of the items by itself (notice that by the symmetry restriction they are priced the same).
	\begin{itemize}
		\item Case $p>1/\varepsilon$:  the buyer buys nothing unless she has valuation $(1/\varepsilon^2, 1/\varepsilon^2)$. Thus the expected revenue is at most $2$.
		\item Case $1< p\leq 1/\varepsilon$: the buyer never buys a single item when his valuation is $(1,1)$. If a buyer with valuation $(1/\varepsilon^2, 1/\varepsilon^2)$ does not buy the bundle, the revenue is bounded by $p\cdot (\varepsilon+\varepsilon^2)\leq (1/\varepsilon) \cdot (\varepsilon+\varepsilon^2) = 1+\varepsilon$. If a buyer with valuation $(1/\varepsilon^2, 1/\varepsilon^2)$ does buy the bundle, it must be the case that $2/\varepsilon^2-q\geq 1/\varepsilon^2 - p$ and so $q\leq 1/\varepsilon^2+p$.
		If $q\leq 1/\varepsilon$ the revenue is bounded from above by $(1/\varepsilon)\cdot (\varepsilon+\varepsilon^2) = 1+\varepsilon$. Otherwise, players with types other than $(1/\varepsilon^2, 1/\varepsilon^2)$ do not buy the bundle, and the revenue is bounded by
		$$ q\cdot \varepsilon^2  + p \cdot \varepsilon  \leq (1/\varepsilon^2+p)\cdot \varepsilon^2  + p\cdot \varepsilon  = 
		1 +p (\varepsilon+\varepsilon^2) \leq 1 + (1/\varepsilon) \cdot (\varepsilon+\varepsilon^2) = 
		2+\varepsilon$$
		\item Case $p\leq 1$: selling items by themselves can yield a revenue of at most $p\cdot 1 \leq 1$ (together). If $q\leq 2$, the revenue is bounded by $2$. If 
		$2<q\leq 1/\varepsilon$ the revenue from selling at $q$ is bounded from above by $(1/\varepsilon)\cdot (\varepsilon+\varepsilon^2) = 1+\varepsilon$, so the total revenue is bounded by $2+\varepsilon$. 
		Otherwise, players with types other than $(1/\varepsilon^2, 1/\varepsilon^2)$ do not buy the bundle, and the revenue is bounded by
		$$ q\cdot \varepsilon^2  + p \cdot 1  \leq (1/\varepsilon^2+p)\cdot \varepsilon^2  + p  = 
		1 +p (1+\varepsilon^2) \leq 1 + 1 \cdot (1+\varepsilon^2) = 
		2+\varepsilon^2$$
	\end{itemize}
	\qed
\end{example}



{

\section{Randomized Mechanisms}\label{sec:app-rand}

For symmetric distributions, the optimal randomized mechanism is trivially symmetric (given an asymmetric mechanism, we can run either this mechanism or its reflection with probability 50\%). Hence Theorem~\ref{thm:symmetric} trivially extends to randomized mechanism.

It is natural to ask whether the result of Theorem~\ref{thm:submodular} can also be extended to randomized mechanisms.
However, it is not even clear how to extend the definition of submodularity to randomized menus. 
Below we discuss a few different natural randomized analogs and observe
that submodular menus need not be optimal for randomized mechanisms in any of these analogs, even for finite-support distributions.

First we note that for two items, the classes of budget-additive, submodular, fractionally subadditive, and subadditive functions are identical (in general there are strict containment between those classes, in the above order). While we don't know of a good analog of submodular functions to randomized mechanisms, the other three notions have natural extensions.

\paragraph{Budget-linear menus}
A {\em budget-additive} menu, charges, for each bundle $S$, the sum of the item prices up to a cap (``budget'') $b$; i.e. $p_S = \min\{b, \sum_{i\in S} p_i\}$.
It is natural to extend such menus to {\em budget-linear} menus where, for each vector $\pi \in [0,1]^n$ of probabilities, the buyer has to pay $$p_{\pi} \triangleq \min\Big\{b, \sum_{i\in [n]} \pi_i \cdot p_i\Big\}.$$

\paragraph{Fractionally subadditive menus}
One way to define fractionally subadditive functions is by their XOS formulation: $p_S = \max_{\alpha\in A} \Big\{\sum_{i \in S} p_{i,\alpha} \Big\}$ (for some finite set $A$). 
Similarly to the budget-linear definition, this can be extended as:
$$ p_{\pi} \triangleq \max_{\alpha} \Big\{\sum_i p_{i,\alpha} \cdot \pi_i \Big\}.$$

\paragraph{False-name-proof mechanisms}
A nice characterization of subadditive deterministic menus is that they are false-name-proof, i.e. a buyer is never better off participating in the mechanism several times.
For randomized mechanisms, there is some subtlety in defining what we mean by participating in the mechanism several times: how do we model the situation where the buyer chooses to receive item $i$ with probabilities $\pi_i$ and then returns choose to receive the same item with probability $\rho_i$?
One option, is to think of the probabilities as fractional allocations of a divisible goods, and let them add up, with a cap at $1$, i.e. $\min\{1, \pi_i + \rho_i\}$. Note that it doesn't make sense to allocate more than $1$ unit of item $i$. Notice also that this definition generalizes budget-linear and fractionally subadditive menus.
Alternatively, returning to the probabilities interpretation, we may think of two independent lotteries, i.e. the new probability is $1-(1-\pi_i)(1-\rho_i)$. For  this option, we again have a couple of modeling choices: either the buyer is adaptive and observes the lottery outcomes before participating in the mechanism again, or non-adaptive.
Notice that the non-adaptive probabilistic buyer is the weakest, hence this is the most general definition of false-name-proofness.

The optimal (randomized) mechanism might not satisfy any of the generalizations of submodular menus suggested above --- even for two items with i.i.d. valuations with finite support. 
We prove that the revenue obtainable from a non-adaptive probabilistic buyer (that is allowed to non-adaptively pick more than one entry from the menu of lotteries) is strictly lower than the revenue obtainable from a buyer that is restricted to pick exactly one entry from the menu.   
\begin{theorem}
	There exist a distribution $\FF$ with finite support such that for two items sampled IID from $\FF$ ($\DD= \FF\times \FF$), the following holds\footnote{Note that since the support is finite, the optimal mechanism in each case is indeed obtained.}:
	When the buyer is allowed to pick only one entry from the menu, there is an optimal randomized mechanism with a finite menu.
	In the model where a non-adaptive probabilistic buyer (see description above) is allowed to pick more than one entry from the menu, the optimum expected revenue is strictly lower. 
\end{theorem} 	
Not surprisingly, the theorem follows directly from an example due to Hart and Reny~\cite[Example 2]{HR15-nonmon}, who used it to show that the optimal mechanism may be non-monotone (recall that submodular mechanisms are always monotone; however the other direction is false).

\begin{example}[Optimal mechanisms are not false-name-proof~\cite{HR15-nonmon}]\label{ex:rand}
	
	Consider the following valuation distribution:
	$$
	\FF \sim \begin{cases}
	10 & \text{w.p. $\frac{2399}{9000}$}\\
	13 & \text{w.p. $\frac{1}{9000}$}\\
	46 & \text{w.p. $\frac{1}{90}$}\\
	47 & \text{w.p. $\frac{1}{3}$}\\
	80 & \text{w.p. $\frac{7}{30}$}\\
	100 & \text{w.p. $\frac{7}{45}$}
	\end{cases}
	$$
	
	Hart and Reny~\cite{HR15-nonmon} argue that the following menu is the unique optimal mechanism for selling two items with i.i.d. valuations, each distributed according to $\FF$:
	\begin{center}
		\renewcommand{\arraystretch}{1.3}
		\begin{tabular}{ c |  c }
			Allocation & Payment\\
			\hline			
			$(0,0)$ & 0 \\
			$\left(\frac{32}{1187}, \frac{384}{13057}\right) $&$ \frac{34240}{13057}$ \\
			$\left(\frac{384}{13057}, \frac{32}{1187}\right) $&$ \frac{34240}{13057}$ \\
			$\left(\frac{35}{1187}, \frac{35}{1187}\right) $&$ \frac{3258}{1187}$ \\
			$\left(\frac{32}{1187}, \frac{5647}{5935}\right) $&$ \frac{90672}{1187}$ \\
			$\left(\frac{5647}{5935}, \frac{32}{1187}\right) $&$ \frac{90672}{1187}$ \\
			$\left(\frac{35}{1187}, \frac{5647}{5935}\right) $&$ \frac{90810}{1187}$ \\
			$\left(\frac{5647}{5935}, \frac{35}{1187}\right) $&$ \frac{90810}{1187}$ \\
			$(0,1)$& $80$\\
			$(1,0)$& $80$ \\
			$(1,1)$& $126$
		\end{tabular}
	\end{center}
	
	In this mechanism, a buyer with valuation $(46, 80)$ chooses bundle $\left(\frac{35}{1187}, \frac{5647}{5935}\right)$ (this can be verified by comparing her utility for each of the 11 menu options); after subtracting her payment, her utility is:
	$$ 46\cdot \frac{35}{1187} + 80 \cdot \frac{5647}{5935} - \frac{90810}{1187} = \frac{1152}{1187} \approx 0.97.$$
	
	In contrast, if allowed to participate in the mechanism twice, the buyer would, for example, prefer the first two menu items, which together give her utility of:
	$$ (46+80)\cdot \Big(1-\Big(1-\frac{32}{1187}\Big)\Big(1-\frac{32}{1187}\Big)\Big) - 2 \cdot \frac{34240}{13057} \approx 1.46.$$
	\qed
\end{example}


In this example, the revenue that the seller obtains from the optimal menu when the buyer is only allowed to pick one entry, is strictly smaller than the revenue from the same menu when the buyer is allowed to pick two entries. We argue that for this example, the optimal revenue from any menu in which the buyer is allowed to pick multiple entries (using ''false names'') is smaller by a positive constant factor from the revenue of the optimal menu when the buyer must pick a single entry. 

Formally, to prove that there is indeed some positive constant 
gap, we have to rule out a sequence of false-name proof mechanisms whose revenue approaches that of Example~\ref{ex:rand}. By~\cite{HR15-nonmon}, the latter is the unique optimum of the standard revenue maximization LP. Furthermore, the set of mechanisms for which the false-name deviation in Example~\ref{ex:rand} does not increase the buyer's expected utility is separated by a linear constraint (hyperplane) from the optimal auction. Therefore, their revenue cannot approach the optimal auction without obtaining it.


Thus, although we do not try to estimate the maximal gap between false-name proof mechanisms and optimal mechanisms, it is interesting 
that the above example proves that the gap is at least some positive constant.

\begin{remark}\label{rem:menu-size}
This is of particular interest in the context of the ``additive-menu-size'' complexity measure suggested in~\cite{HartN13}. 
\citet{HartN13} observe that even simple mechanisms like selling each item separately have exponentially long description ({\em menu-size}) when written as a menu, or list of (allocation, price) options presented to the buyer.
To alleviate this issue, they proposed a relaxed measure of {\em additive-menu-size} where the buyer can select multiple options from a menu; for example, selling each item separately has only a linear additive-menu-size.
The former observation was used in~\cite{BGN17-menu-complexity} to prove an exponential lower bound on the menu-size complexity of any mechanism that achieves near-optimal revenue; a prelimenary version of~\cite{BGN17-menu-complexity} left as an open problem whether better upper bounds on additive-menu-size of near-optimal mechanism can be obtained in general. 
Example~\ref{ex:rand} in this section answers this question on the negative in a very strong sense: the example implies that for some distributions additive-menu-size complexity is not even defined for any near-optimal mechanism.

\end{remark}


\ifFULL
\section*{Acknowledgments}
This project has received funding from the European Research Council (ERC) under the European Union’s
Horizon 2020 research and innovation programme (grant agreement No 740282).

A.R.'s work was mostly done while he was an intern at Microsoft Research, and a visitor at Hebrew University (supported by above ERC grant). Parts of this work were also done while at Harvard University (supported by the Rabin Postdoctoral Fellowship).
\bibliographystyle{alpha}
\else
\bibliographystyle{ACM-Reference-Format}
\fi
\bibliography{drev}

\ifFULL
\else
\appendix

\section{Proof of Theorem~\ref{thm:intro-symmetric}}
\label{app:iid}

\fi	 			
\end{document}